\documentclass[english,11pt]{article}
\usepackage[utf8]{inputenc}
\usepackage{amsfonts,amssymb,verbatim,bm,latexsym,amsmath,url,amsbsy,natbib,float,placeins,flafter} 
\usepackage[pdftex]{graphicx}
\usepackage{lscape, booktabs, accents,a4wide}
\usepackage[usenames]{color}

\numberwithin{equation}{section}
\newtheorem{theorem}{Theorem}[section]
\newtheorem{assumption}{Assumption}[section]

\newtheorem{lemma}{Lemma}[section]
\newtheorem{remark}{Remark}[section]

\newcommand\independent{\protect\mathpalette{\protect\independenT}{\perp}}
\def\independenT#1#2{\mathrel{\rlap{$#1#2$}\mkern2mu{#1#2}}}

\newcommand{\ubar}[1]{\underaccent{\bar}{#1}}

\newcommand\norm[1]{\left\lVert#1\right\rVert}
\newcommand{\E}{\mathbb{E}}
\newcommand{\R}{\mathbb{R}}

\begin{document}

\title{An alternative to synthetic control for models with many covariates under sparsity\thanks{We thank participants at the 2016 North American and European Meetings of the Econometric Society, the 2017 IAAE Meeting and CREST internal seminars for their useful comments and discussions. We acknowledge funding from Investissements d'Avenir (ANR-11-IDEX-0003/Labex Ecodec/ANR-11-LABX-0047).}}

\author{Marianne~Bl\'{e}haut\thanks{CREST-ENSAE, marianne.blehaut@ensae.fr.} \and Xavier~D'Haultf\oe{}uille\thanks{CREST-ENSAE, xavier.dhaultfoeuille@ensae.fr} \and J\'{e}r\'{e}my~L'Hour\thanks{CREST-ENSAE, jeremy.l.hour@ensae.fr.} \and Alexandre~B.~Tsybakov\thanks{CREST-ENSAE, alexandre.tsybakov@ensae.fr.}}

\maketitle

\begin{abstract}
The synthetic control method is a an econometric tool to evaluate causal effects when only one unit is treated. While initially aimed at evaluating the effect of large-scale macroeconomic changes with very few available control units, it has increasingly been used in place of more well-known microeconometric tools in a broad range of applications, but its properties in this context are unknown. This paper introduces an alternative to the synthetic control method, which is developed both in the usual asymptotic framework and in the high-dimensional scenario.  We propose an estimator of average treatment effect that is doubly robust, consistent and asymptotically normal. It is also immunized against first-step selection mistakes. We illustrate these properties using Monte Carlo simulations and applications to both standard and potentially high-dimensional settings, and offer a comparison with the synthetic control method.
        
\end{abstract}

\section{Introduction}

The synthetic control method \citep{AbadieGardeazabal2003,AbadieDiamondHainmueller2010,AbadieDiamondHeinmueller2015} is one of the most recent additions to the empiricist's toolbox, gaining popularity not only in economics, but also in political science, medicine, etc. It provides a sound methodology in many settings where only long aggregate panel data is available to the researcher. The method has been specifically developed in a context where a single sizeable unit such as a country, a state or a city undergoes a large-scale policy change (referred to as the treatment or intervention hereafter), while only a moderate number of control units (the donor pool) is available to construct a counterfactual through a synthetic unit. This unit is defined as a convex combination of units from the donor pool that best resembles the treated unit before the intervention. Then the treatment effect is estimated from the difference in outcomes between the treated unit and its synthetic unit after the intervention takes place. In contexts such as those described above, the synthetic unit possesses several appealing properties \citep[for more details on such properties, see the recent survey of][]{abadie2019using}. First, it does not lead to extrapolation outside the support of the data: because weights are non-negative, the counterfactual never takes a value outside of the convex hull defined by the donor pool. Second, one can assess simply its fit, making it easy to judge the quality of the counterfactual. Third, the  synthetic unit is sparse: the number of control units receiving a non-zero weight is at most equal to the dimension of the matching variable plus one.
 
\medskip
The method has still some limitations, in particular when applied to micro data, for which it was not initially intended. In such cases, the number of untreated units $n_0$ is typically greater than the dimension $p$ of variables $X$ used to construct the synthetic units. Then, as soon as the treated unit falls into the convex hull defined by the donor pool, the synthetic control solution is not uniquely defined  \citep[see in particular][]{abadie2017penalized}. Second, and still related to the fact that the method was not developed for micro data, there is, to the best of our knowledge, no asymptotic theory available for synthetic control yet. This means in particular, that inference cannot be conducted in a standard way. A third issue is related to variable selection. The standard synthetic control method, as advocated in \cite{AbadieDiamondHainmueller2010}, not only minimizes the norm $\norm{.}_V$ -- defined for a vector $a$ of dimension $p$ and diagonal positive-definite matrix $V$, as $\norm{a}_V = \sqrt{a^T V a}$ -- between the characteristics of the treated and those of its synthetic unit under constraints, but also employs a bi-level optimization program over the weighting matrix $V$ so as to obtain the best possible pre-treatment fit. Diagonal elements of $V$ are interpreted as a measure of the predicting power of each characteristics for the outcome  \citep[see, e.g.,][]{AbadieDiamondHainmueller2010,abadie2019using}. This approach has been criticized for being unstable and yielding unreproducible results, see in particular \cite{klossner2018comparative}.

\medskip
We consider an alternative to the synthetic control that addresses these issues. Specifically, we consider a parametric form for the synthetic control weights, $W_i = h(X_i^T\beta_0)$, where we estimate the unknown parameter $\beta_0$. This approach warrants the uniqueness of the solution in low-dimensional cases where $p<n_0$. With micro data, it may thus be seen as a particular solution of the synthetic control method. We show that the average treatment on the treated (ATT) parameter can be estimated with a two-step GMM estimator, where $\beta_0$ is computed in a first step so that the reweighted control group matches some features of the treated units. A key result  is the double robustness of the estimator, as defined by \cite{RobinsBang2005}. Specifically, we show that misspecifications in the synthetic control weights do not prevent valid inference if the outcome regression function is linear for the control group. 

\medskip
We then turn to the high-dimensional case where $p$ is large, possibly greater than $n_0$. This case actually corresponds to the initial set-up of the synthetic control method, and is therefore crucial to take into consideration. We depart from the synthetic control method by introducing an $\ell_1$ penalization term in the minimization program used to estimate $\beta_0$. We thus perform variable selection in a similar way as the Lasso, but differently from the synthetic control method, which relies on the aforementioned optimization over $V$ (leading to overweighting the variables that are good predictors of the outcome and underweighting the others).

\medskip
We also study the asymptotic properties of our estimator. Building on double robustness, we construct an estimator that is immunized against first-step selection mistakes in the sense defined for example by \cite{ChernozhukovHansenSpindler2015b} or \cite{DoubleML2018}. This construction requires an extra step, which models the outcome regression function and provides a bias correction, a theme that has also been developed in \cite{ben2018augmented}, \cite{abadie2017penalized} and \cite{SyntheticDiffinDiff2019}. We show that both in the low- and high-dimensional case, the estimator is consistent and asymptotically normal. 
Consequently, we develop inference based on asymptotic approximation, which can be used in place of permutation tests when randomization of the treatment is not warranted.

\medskip
Apart from its close connection with the synthetic control method, the present paper is related to the literature on treatment effect evaluation through propensity score weighting and covariate balancing. Several efforts have been made to include balance between covariates as an explicit objective for estimation with or without relation to the propensity score \citep[e.g.][]{Hainmueller2012, Graham2012}. Our paper is in particular related to that of \cite{ImaiRatkovic2014}, who integrate propensity score estimation and covariate balancing in the same framework. We extend their paper by considering the case of high-dimensional covariates. Note that the covariate balancing idea is related to the calibration estimation in survey sampling, see in particular \cite{deville1992calibration}.

\medskip
It also partakes in the econometric literature addressing variable selection, and more generally the use of machine learning tools, when estimating a treatment effect, especially but not exclusively in a high-dimensional framework. The lack of uniformity for inference after a selection step has been raised in a series of papers by \cite{LeebPotscher2005,LeebPotscher2008a,LeebPotscher2008b}, echoing earlier papers by \cite{Leamer1983} who put into question the credibility of many empirical policy evaluation results. One recent solution proposed to circumvent this post-selection conundrum is the use of double-selection procedures \citep{BelloniChernozhukov2013,Farrell2015,ChernozhukovHansenSpindler2015,DoubleML2018}. For example, \cite{BCH2012Treat} highlight the dangers of selecting controls exclusively in their relation to the outcome and propose a three-step procedure that helps selecting more controls and guards against omitted variable biases much more than a simple ``post-single-selection'' estimator, as it is usually done by selecting covariates based on either their relation with the outcome or with the treatment variable, but rarely both. \cite{Farrell2015} extends the main approach of \cite{BCH2012Treat} by allowing for heterogeneous treatment effects, proposing an estimator that is robust to either model selection mistakes in propensity scores or in outcome regression. However, \cite{Farrell2015} proves the root-n consistency of his estimator only when both models hold true (see Assumption 3 and Theorem 3 therein). Our paper is also related to the work of \cite{athey2018}, who consider treatment effect estimation under the assumption of a linear conditional expectation for the outcome equation. As we do, they also estimate balancing weights, to correct for the bias arising in this high-dimensional setting, but because of their linearity assumption, they do not require to estimate a propensity score. Their method is then somewhat simpler than ours, but it does not enjoy the double-robustness property of ours. 

\medskip
Finally, a recent work by \cite{Wager2019} parallel to ours\footnote{Our results have been presented as early as 2016 at the North American and European Summer Meetings of the Econometric Society (see \url{ https://www.econometricsociety.org/sites/default/files/regions/program_ESEM2016.pdf}), and again in 2017 during the IAAE Meeting in Sapporo, see \cite{IAAE2017Sapporo}.} suggests a Lasso-type procedure assuming logistic propensity score and linear specification for both treated and untreated items.  Their  method is similar but still slightly different from ours. The main focus in \cite{Wager2019} is to prove asymptotic normality under possibly weaker conditions on the sparsity levels of the parameters. Namely, they allow for sparsity up to  $o(\sqrt{n} /\log(p))$ or, in some cases up to $o(n /\log(p))$ where~$n$ is the sample size, when the eigenvalues of the population Gram matrix do not depend on~$n$. Similar results can be developed for our method at the expense of additional technical effort. We have opted not to pursue in this direction since it only helps to include relatively non-sparse models that are not of interest for the applications we have in mind. More recently, the papers of \cite{NingImaiSida2018} and \cite{tan2020} written independently have been brought to our attention. \cite{tan2020} studies double-robustness of a similar estimator but assumes a logistic propensity score model while we leave room for other possibilities. The setting in \cite{NingImaiSida2018} is not restricted to the logistic model. However, that paper  considers propensity score only as a mean to select the variables that should be fully balanced and that will enter in the final propensity score. This leads to a more complex estimation procedure. The authors of these two papers do not seem to be aware of the prior work of ours \citep{IAAE2017Sapporo}.

\medskip
The paper is organized as follows. Section \ref{sec:setup} introduces the set-up and the identification strategy behind our estimator. Section \ref{sec:EstimStrategy} presents the estimator both in the low- and high-dimensional case and studies its asymptotic properties. Section \ref{sec:MCXP} examines the finite sample properties of our estimator through a Monte Carlo experiment. Section \ref{sec:applications} revisits \cite{lal86}'s dataset to compare our procedure with other high-dimensional econometric tools and the effect of the large-scale tobacco control program of \cite{AbadieDiamondHainmueller2010} for a comparison with synthetic control. Section \ref{sec:conclusion} concludes. All proofs are in the Appendix. 

\section{Covariate Balancing Weights and Double Robustness}
\label{sec:setup}

We are interested in the effect of a binary treatment, coded by $D=1$ for the treated and $D=0$ for the non-treated. We let $Y(0)$ and $Y(1)$ denote  the potential outcome under no treatment and under the treatment, respectively. The observed outcome is then $Y=D Y(1) + (1-D) Y(0)$. We also observe a random vector $X \in \mathbb{R}^p$ of pre-treatment characteristics. 
The quantity of interest is the average treatment effect on the treated (ATT), defined as:
\begin{equation*}
\theta_0 = \mathbb{E}[Y(1)-Y(0) \vert D = 1].
\end{equation*}
Here and in what follows, we assume that the random variables  are such that all the considered expectations are finite.  Since no individual is observed in both treatment states, identification of the counterfactual $\mathbb{E}[Y(0) \vert D=1]$ is achieved through the following two ubiquitous conditions.

\begin{assumption}[Nested Support] \label{ass:nonmadattreat}
$P[D=1 \vert X] < 1$ almost surely and $0<P[D=1] <1$.
\end{assumption}

\begin{assumption}[Mean Independence] \label{ass:meanIndep}
$\mathbb{E}[Y(0)\vert X, D=1] = \mathbb{E}[Y(0)\vert X, D=0]$.
\end{assumption}

Assumption \ref{ass:nonmadattreat}, a version of the usual common support condition, requires that there exist control units for any possible value of the covariates in the population. Since the ATT is the parameter of interest, we are never reconstructing a counterfactual for control units so $P[D=1 \vert X] > 0$ is not required. Assumption \ref{ass:meanIndep} states that conditional on a set of observed confounding factors, the expected potential outcome under no treatment is the same for treated and control individuals. This assumption is a weaker form of the classical conditional independence assumption $(Y(0), Y(1)) \independent D \vert X$. 
 
\medskip
In policy evaluation settings, the counterfactual is usually identified and estimated as a weighted average of non-treated unit outcomes:
\begin{equation} \label{eq:ATTWeighted}
\theta_0 = \mathbb{E}[Y(1) \vert D = 1] - \mathbb{E}[WY(0) \vert D = 0],
\end{equation}
where $W$ is a random variable called the weight. Popular choices for the weight are the following:
\begin{enumerate}
\item Linear regression: $W= \mathbb{E}[DX^T] \mathbb{E}[(1-D) X X^T]^{-1}X$, also referred to as the Oaxaca-Blinder weight \citep{Kline2011};
\item Propensity score: $W=P[D=1\vert X] / (1-P[D=1\vert X])$;
\item Matching: $W=P(D=1)f_{X|D=1}(X)/[P(D=0)f_{X|D=0}(X)]$;
\footnote{Assuming here that the conditional densities $f_{X|D=1}$ and $f_{X|D=0}$ exist.  Of course, $P(D=1)f_{X|D=1}(X)/[P(D=0)f_{X|D=0}(X)]=P[D=1\vert X] / (1-P[D=1\vert X])$, but the methods of estimation of $W$ differ in the two cases.}
\item Synthetic controls: see \cite{AbadieDiamondHainmueller2010}.
\end{enumerate}

In this paper, we propose another choice of weight $W$, which can be seen as a  parametric alternative to the synthetic control. An advantage is that it is well-defined whether or not  the number of untreated observations $n_0$ is greater than  the dimension $p$ of $X$, whereas the synthetic control estimator is not uniquely defined when $n_0>p$. Formally, we look for weights $W$ that:
\begin{itemize}
\item[(i)] satisfy a balancing condition as in the synthetic control method;
\item[(ii)] are positive;
\item[(iii)] depend only on the covariates;
\item[(iv)] can be used whether $n_0>p$ or $n_0 \leq p$ (high-dimensional regime).
\end{itemize}
Satisfying a balancing condition means that
\begin{equation} \label{BalancingCondition}
 \mathbb{E}[DX] = \mathbb{E}[W(1-D)X].
\end{equation}
Up to a proportionality constant, this is equivalent to $\mathbb{E}[X \vert D=1] = \mathbb{E}[WX \vert D=0]$. In words,~$W$ balances the first moment of the observed covariates between the treated and the control group. The definition of the observable covariates $X$ is left to the econometrician and can include transformation of the original covariates so as to match more features of their distribution. The idea behind such weights relies on the principle of ``covariate balancing'' as in, e.g., \cite{ImaiRatkovic2014}. The following lemma shows that under Assumption  \ref{ass:nonmadattreat} weights satisfying the balancing condition always exist. 

\begin{lemma}\label{lem:2.1}
If Assumption \ref{ass:nonmadattreat} holds, the propensity score weight $W= P[D=1 \vert X]/(1-P[D=1 \vert X])$ satisfies the balancing condition \eqref{BalancingCondition}.
\end{lemma}

The proof of this lemma is straightforward by plugging the expression of $W$ in Equation \eqref{BalancingCondition} and using the law of iterated expectations. Note that $W$ is not a unique solution of \eqref{BalancingCondition}. The linear regression weight $W= \mathbb{E}[DX^T] \mathbb{E}[(1-D) X X^T]^{-1}X$ also satisfies the balancing condition but it can be negative and its use is problematic in high-dimensional regime.  

\medskip
Lemma \ref{lem:2.1} would suggest solving a binary choice model to obtain estimators of $P[D=1 \vert X]$ and of the weight~$W$ as a first step, and then plugging $W$ in \eqref{eq:ATTWeighted} to estimate $\theta_0$. However, an inconsistent estimator of the propensity score leads to an inconsistent estimator of $\theta_0$ and does not guarantee that the corresponding weight will achieve covariate balancing. Finally, estimation of the propensity score can be problematic when there are very few treated units. For these reasons, we consider another approach where estimation is based directly on balancing equations:
\begin{equation}
    \mathbb{E}\left[ \left(D - (1-D) W\right) X\right]=0.
    \label{eq:calibr}
\end{equation}
An important advantage of this approach over the usual one based on the propensity score estimation through maximum likelihood is its double robustness \cite[for a definition, see, e.g.,][]{RobinsBang2005}. Indeed, let $W_1$ denote the weights identified by \eqref{eq:calibr} under a misspecified model on the propensity score. It turns out  that if the balancing equations $\eqref{eq:calibr}$ hold for $W_1$ the estimated treatment effect will still be consistent provided that $\mathbb{E}[Y(0)|X]$ is linear in $X$. The formal result is given in Theorem \ref{thm:double_rob0} below.

\begin{theorem}[Double Robustness]
\label{thm:double_rob0}
Let Assumptions \ref{ass:nonmadattreat}-\ref{ass:meanIndep} hold and let $w:\mathbb{R}^p\to (0,+\infty)$ be a  measurable function such that $\mathbb{E}\left[ w(X)|Y|\right] <\infty$, $\mathbb{E}\left[ w(X)\| X\|_2\right] <\infty$, where $\|\cdot\|_2$ denotes the Euclidean norm. Assume the
balancing condition 
\begin{equation}
    \mathbb{E}\left[ \left(D - (1-D) w(X)\right) X\right]=0.
    \label{eq:calibr0}
\end{equation}
Then, for any $\mu \in \mathbb{R}^p$ the ATT \ $\theta_0$ can be expressed as
\begin{equation} \label{ATTEstimMoment0}
\theta_0 = \frac{1}{P(D=1)} \mathbb{E}\left[ \left(D - (1-D)w(X) \right) (Y-X^T\mu)\right]
\end{equation}
in each of the following two cases:
\begin{enumerate}
\item $\mathbb{E} [Y(0) \vert X] = X^T \mu_0$ for some $\mu_0 \in \mathbb{R}^p$;
\item $P[D=1|X]=w(X)/(1+w(X))$.
\end{enumerate}
\end{theorem}
In \eqref{ATTEstimMoment0}, the effect of $X$ is taken out from $Y$ in a linear fashion, while the effect of $X$ on $D$ is taken out by re-weighting the control group to obtain the same mean for $X$.  Theorem \ref{thm:double_rob0} shows that an estimator based on  \eqref{ATTEstimMoment0} enjoys the double robustness property. Theorem \ref{thm:double_rob0} is similar to the result of \cite{Kline2011} for the Oaxaca-Blinder estimator, which is obtained under the assumption that the propensity score follows specifically a log-logistic model in the propensity-score-well-specified case. Theorem \ref{thm:double_rob0} is more general. It can be applied under parametric modeling of  $W$ as well as in nonparametric settings. 

\medskip
In this paper, we consider a parametric model for $w(X)$. Namely, we assume that $P[D=1|X]=G(X^T\beta_0)$ for some unknown $\beta_0\in \mathbb{R}^p$ and some known strictly increasing cumulative distribution function $G$. Then $w(X)=h(X^T\beta_0)$ with $h = G/(1-G)$ and $\beta_0$ is identified by the balancing condition
\begin{equation} \label{eq:calibr-our}
    \mathbb{E}\left[ \left(D - (1-D) h(X^T\beta_0)\right) X\right]=0.
   \end{equation}
Clearly, $h$ is a positive strictly increasing function, which implies that its primitive $H$ is strictly convex. A classical example is to take $G$ as the c.d.f. of the logistic distribution, in which case  $h(u)= H(u)=\exp(u)$ for $u\in \mathbb{R}$. The strict convexity of $H$ implies that 
$\beta_0$ is the unique solution of a strictly convex program: 
	\begin{equation} \label{eq:convex_prgm}
		\beta_0=\underset{\beta\in \mathbb{R}^p}{\arg \min} \; \mathbb{E} \left[(1-D)H(X^T\beta) - D X^T\beta\right].	
	\end{equation}
This program is well-defined, whether or not $P[D=1|X]=G(X^T\beta_0)$. Note also that definitions \eqref{eq:calibr-our} and \eqref{eq:convex_prgm} are equivalent provided that $\mathbb{E}\left[ h(X^T\beta)\| X\|_2\right] <\infty$ for $\beta$ in a vicinity of $\beta_0$. Indeed,  it follows from the dominated convergence theorem that, under this assumption and due to the fact that any convex function is locally Lipschitz, differentiation under the expectation sign is legitimate in \eqref{eq:convex_prgm}.

\medskip
We are now ready to state the main identification theorem justifying the use of ATT estimation methods developed below. It is a straightforward corollary of Theorem~\ref{thm:double_rob0}.
\begin{theorem}[Parametric Double Robustness]
\label{thm:double_rob_param}
Let Assumptions \ref{ass:nonmadattreat}-\ref{ass:meanIndep} hold. Assume that $\beta_0\in \mathbb{R}^p$ and a positive strictly increasing function $h$ are such that $\mathbb{E}\left[ h(X^T\beta_0)\| X\|_2\right] <\infty$,  $\mathbb{E}\left[ h(X^T\beta_0)| Y|\right] <\infty$ and condition \eqref{eq:calibr-our} holds. Then, for any $\mu \in \mathbb{R}^p$, the ATT \ $\theta_0$ satisfies
\begin{equation} \label{ATTEstimMoment}
\theta_0 = \frac{1}{P(D=1)} \mathbb{E}\left[ \left(D - (1-D)h(X^T \beta_0) \right) (Y-X^T\mu)\right],
\end{equation}
in each of two the following cases.
\begin{enumerate}
\item 
There exists $\mu_0 \in \mathbb{R}^p$ such that $\mathbb{E} [Y(0) \vert X] = X^T \mu_0$.
\item $P[D=1|X]=G(X^T\beta_0)$ with $G=h/(1+h)$.
\end{enumerate}
\end{theorem}
At this stage, the parameter $\mu$ in Equation \eqref{ATTEstimMoment} does not play any role and can, for example, be zero. However, we will see below that in the high-dimensional regime,  choosing  $\mu$ carefully is crucial  to obtain an ``immunized'' estimator of $\theta_0$ that enjoys the desirable asymptotic properties.


\section{A Parametric Alternative to Synthetic Control}
\label{sec:EstimStrategy}

We now assume to have a sample $(D_i,X_i,Y_i)_{i=1...n}$ of i.i.d. random variables with the same distribution as $(D,X,Y)$.

\subsection{Estimation with low-dimensional covariates}\label{sec:low}

Consider first an asymptotic regime where the dimension $p$ of the covariates is fixed, while the sample size $n$ tends to infinity. We call it the low-dimensional regime. Define an estimator of $\beta_0$ via the empirical counterpart of \eqref{eq:convex_prgm}:
\begin{equation} \label{eq:betaGMM}
\hat \beta_{\rm ld} \in \underset{\beta\in \mathbb{R}^p}{\arg \min} \; \frac{1}{n} \sum_{i=1}^n [(1-D_i) H(X_i^T\beta) - D_i  X_i^T \beta].
\end{equation}
Next, we plug $\hat\beta_{\rm ld}$  in the empirical counterpart of \eqref{ATTEstimMoment} to obtain the following estimator of $\theta_0$:
\begin{equation*}
	     \widetilde{\theta}_{\rm ld} := \frac{1}{ \frac{1}{n} \sum_{i=1}^n D_i} \left(\frac{1}{n}\sum_{i=1}^n [D_i - (1-D_i) h(X_i^T\hat{\beta}_{\rm ld})] Y_i \right).
\end{equation*}
Note that if $X$ includes the intercept, $\widetilde{\theta}_{\rm ld}$ satisfies the desirable property of location invariance, namely it does not change if we replace all $Y_i$ by  $Y_i+c$, for any $c\in\R$.

\medskip
Set $Z:=(D,X,Y)$, $Z_i := (D_i,X_i,Y_i)$ and introduce the function
$$
g(Z,\theta,(\beta,\mu)):= [D - (1-D) h(X^T \beta) ] [Y - X^T \mu] - D \theta.
$$
Then the estimator $\widetilde \theta_{\rm ld}$ satisfies 
\begin{equation}
\frac{1}{n}\sum_{i=1}^n g(Z_i, \widetilde \theta_{\rm ld}, (\hat \beta_{\rm ld},0)) =0.	
	\label{eq:theta_ld}
\end{equation}
This estimator is a two-step GMM. It is consistent and asymptotically normal under mild regularity conditions, with asymptotic variance $\mathbb{E} \left[ g^2(Z, \theta_0, (\beta_0, \mu_0))\right]/\mathbb{E}(D)^2$, where 
\begin{equation}\label{lowmu}
\mu_0= \mathbb{E}[h'(X^T \beta_0) X X^T\vert D=0]^{-1} \mathbb{E}[h'(X^T \beta_0)X Y \vert D=0].
\end{equation}
This can be shown by standard techniques \citep[see, e.g., Section 6 in][]{NeweyMcFadden1994}. Notice that since $h'(X^T \beta_0)>0$ the vector $\mu_0$ is the coefficient of the weighted population regression of $Y$ on $X$ for the control group. This observation is useful for the derivation of the ``immunized'' estimator in the high-dimensional case, to which we now turn.

\subsection{Estimation with high-dimensional covariates}

We now consider that $p$ may grow with $n$, with possibly $p\gg n$. This can be of interest in several situations. First, in macroeconomic problems, $n$ is actually small, and $p$ may easily be of comparable size. For example, in the Tobacco control program application by \cite{AbadieDiamondHainmueller2010} the control group size is limited due to the fact that the observational unit is the state but many pre-treatment outcomes are included among the covariates. Section \ref{sub:tobacco} revisits this example. Second, researcher may want to consider a flexible form for the weights by including transformations of the covariates. For instance, one may want to interact categorical variables with other covariates or consider, e.g., different powers of continuous variables if one wants to allow for flexible non-linear effects. See Section \ref{sub:NSW} for an application considering such transformations. Third, one may want not only to balance the first moments of the distribution of the covariates but also the second moments, the covariances, the third moments and so on to make the distribution more similar between the treated and the control group. In this case, high-dimensional settings seem to be of interest as well.

\medskip
In high-dimensional regime, the GMM estimator in \eqref{eq:betaGMM} is, in general, not consistent.  We therefore propose an alternative Lasso-type method by adding in \eqref{eq:betaGMM} an $\ell_1$ penalization term: 
\begin{equation} \label{betaLasso}
\hat \beta \in\underset{\beta\in \mathbb{R}^p}{\arg \min} \; \left(\frac{1}{n} \sum_{i=1}^n [(1-D_i) H(X_i^T\beta) - D_i X_i^T \beta] + \lambda \sum_{j=1}^{p} \psi_{j} \vert \beta_j \vert\right).
\end{equation}
Here, $\lambda > 0$ is an overall penalty parameter set to dominate the noise in the gradient of the objective function and $\left\{ \psi_{j} \right\}_{j=1, \ldots,p}$ are covariate specific penalty loadings set as to grant good asymptotic properties. The penalty loadings can be adjusted  using the algorithm presented in Appendix \ref{alg}. 

\medskip
This type of penalization offers several advantages. First, the program \eqref{betaLasso} has almost surely a unique solution when the entries of $X$ have a continuous distribution, cf. Lemma 5 in \cite{Tibshirani2013}, which cannot be granted for its non-penalized version \eqref{eq:betaGMM}. Second, it yields a sparse solution in the sense that some entries of the vector of estimated coefficients are set exactly to zero if the penalty is large enough, which is not the case for estimators based on $\ell_2$ penalization. The $\ell_0$-penalized estimator shares the same sparsity property but is very costly to compute, whereas \eqref{betaLasso} can be easily solved by computationally efficient methods, see, e.g., \cite{StatLearning}. 

\medskip
The use of covariate specific penalty loadings goes back to \cite{BRT}; the particular choice of penalty loadings that we consider below is inspired by~\cite{BelloniChenChernozhukovHansen2012}. A drawback of penalizing by the $\ell_1$-norm is that it induces a bias in estimation of the coefficients. But this is not an issue here since we are ultimately interested in estimating $\theta_0$ rather than $\beta_0$. The solution $\hat\beta$ of \eqref{betaLasso} only plays the role of a pilot estimator.

\medskip
The estimator $\hat\beta$ is consistent as $n$ tends to infinity under assumptions analogous to those used in \cite{BTW, BRT} for the Lasso with quadratic loss, see Theorem \ref{the:nuisparam} below. As in the low-dimensional case (cf. Section~\ref{sec:low}), one is then tempted to consider the plug-in estimator for the ATT based on Equation \eqref{ATTEstimMoment} with $\mu=0$:
\begin{equation} \label{eq:naiveplugin}
	     \widetilde{\theta} = \frac{1}{\sum_{i=1}^n D_i} \sum_{i=1}^n [D_i - (1-D_i) h(X_i^T\hat{\beta})] Y_i.
\end{equation}
 We refer to this estimator as the naive plug-in estimator. 
 However, as mentioned above, the Lasso estimator $\hat{\beta}$ of the nuisance parameter $\beta_0$ is not asymptotically unbiased. In high-dimensional regime where $p$ grows with $n$, naive plug-in estimators suffer from a regularization bias and may not be asymptotically normal with zero mean, as illustrated for example in \cite{BCH2012Treat,ChernozhukovHansenSpindler2015b,DoubleML2018}.  Therefore, following the general approach of \cite{ChernozhukovHansenSpindler2015b,DoubleML2018}, we develop an immunized estimator that, at the first order, is insensitive to $\hat{\beta}$. We show that this estimator is asymptotically normal with mean zero and an asymptotic variance that does not depend on the properties of the pilot estimator $\hat{\beta}$. The idea is to choose parameter $\mu$ in \eqref{ATTEstimMoment} such that the expected gradient of the estimating function $g(Z,\theta,(\beta,\mu))$ with respect to $\beta$ is zero when taken at $(\theta_0,\beta_0)$. This holds for $\mu=\mu_0$, where $\mu_0$ satisfies
\begin{equation} \label{eq:mu0_OLS0}
    \mathbb{E} \left[ (1-D) h'(X^T \beta_0) (Y - X^T \mu_0) X \right] = 0.
\end{equation}
Notice that if the corresponding matrix is invertible we get the low-dimensional solution \eqref{lowmu}. Clearly, $\mu_0$ depends on unknown quantities and we need to estimate it.  To this end, observe that Equation \eqref{eq:mu0_OLS0} corresponds to the first-order condition of a weighted least-squares problem,\footnote{The assumptions under which we prove the results below guarantee that $\mu_0$ defined here is unique. Extension to the case of multiple solutions can be worked out as well. It is technically more involved  but in our opinion does not add much to the understanding of the problem.} namely 
\begin{equation}
  \mu_0 = \underset{\mu\in \mathbb{R}^p}{\arg \min} \; \mathbb{E} \left[ (1-D) h'(X^T \beta_0)  (Y - X^T\mu)^2 \right].
  \label{eq:mu0_OLS}
\end{equation}
Since $X$ is high-dimensional we cannot estimate $\mu_0$ via the empirical counterpart of \eqref{eq:mu0_OLS}. Instead, we consider a Lasso-type estimator 
\begin{equation} \label{muLasso}
  \hat \mu \in \underset{\mu\in \mathbb{R}^p}{\arg \min}   \left(\frac{1}{n} \sum_{i=1}^n [(1-D_i) h'(X_i^T \hat \beta )  \left(Y_i - X_i^T \mu\right)^2] + \lambda' \sum_{j=1}^p \psi'_{j} \vert \mu_j \vert\right).
\end{equation}
Here,  similarly to  \eqref{betaLasso}, $\lambda' > 0$ is an overall penalty parameter set to dominate the noise in the gradient of the objective function and $\left\{ \psi'_{j} \right\}_{j=1, \ldots, p}$ are covariate-specific penalty loadings. Importantly, by estimating $\mu_0$ we do not introduce, at least asymptotically, an additional source of variability since by construction, the gradient of the moment condition \eqref{ATTEstimMoment} (if we consider it as function of $\beta$ rather than of $\beta_0$) with respect to $(\beta,\mu)$ vanishes at point $(\beta_0,\mu_0)$.

\medskip
Finally, the immunized ATT estimator is defined as
	\begin{align*}
	    \hat{\theta} &:= \frac{1}{\sum_{i=1}^n D_i} \sum_{i=1}^n \left(D_i-(1-D_i)h(X_i^T \hat \beta)\right)(Y_i-X_i^T \hat \mu ).
	\end{align*}
Intuitively, the immunized procedure corrects the naive plug-in estimator in the case where the balancing program has ``missed'' a covariate that is very important to predict the outcome: 
$$\hat{\theta}= \tilde \theta - 
\left[\frac{1}{n_1} \sum_{i: D_i = 1}^n X_i -\frac{1}{n_1} \sum_{i: D_i = 0}^n h(X_i^T\hat \beta)X_i \right]^T \hat \mu,
$$
where $n_1$ is the number of treated observations.
This has a flavor of Frish-Waugh-Lowell partialling-out procedure for model selection as observed by \cite{BCH2012Treat} and further developed in \cite{ChernozhukovHansenSpindler2015b}.

\medskip
To summarize, the estimation procedure in high-dimensional regime consists of the three following steps. Each step is computationally simple as it needs at most to minimize a convex function:

\begin{enumerate}
\item \textit{(Balancing step.)} For a given level of penalty $\lambda$ and positive covariate-specific penalty loadings $\left\{ \psi_{j} \right\}^{p}_{j=1}$, compute $\hat\beta$ defined by
\begin{equation}
\hat \beta \in \underset{\beta\in \mathbb{R}^p}{\arg \min} \;\left(\frac{1}{n} \sum_{i=1}^n [(1-D_i) H(X_i^T\beta) - D_i X_i^T \beta] + \lambda \sum_{j=1}^{p} \psi_{j} \vert \beta_j \vert\right).
\end{equation}
\item \textit{(Immunization step.)} For a given level of penalty $\lambda'$ and covariate-specific penalty loadings $\left\{ \psi'_{j} \right\}^{p}_{j=1}$, and using $\hat \beta$ obtained in the previous step, compute $\hat\mu$ defined by:
\begin{equation}
  \hat \mu \in \underset{\mu\in \mathbb{R}^p}{\arg \min}   \left(\frac{1}{n} \sum_{i=1}^n  [(1-D_i) h'(X_i^T \hat \beta)  \left(Y_i - X_i^T \mu\right)^2]  + \lambda' \sum_{j=1}^p \psi'_{j} \vert \mu_j \vert\right).
\end{equation}
\item \textit{(ATT estimation.)} Estimate the ATT using the immunized estimator:
	\begin{equation} \label{eq:thetaimmun}
	    \hat{\theta} = \frac{1}{\sum_{i=1}^n D_i} \sum_{i=1}^n \left[D_i-(1-D_i)h(X_i^T \hat \beta)\right](Y_i-X_i^T \hat \mu ).
	\end{equation}
\end{enumerate}

\subsection{Asymptotic Properties}
\label{sec:Asymptotics}

The current framework poses several challenges to achieving asymptotically valid inference. 
First, $X$ can be high-dimensional since we allow for $p \gg n$ provided that sparsity conditions are met (see Assumption \ref{ass:dimres} below). Second, the ATT estimation is affected by the estimation of the nuisance parameters $(\beta_0,\mu_0)$ and we wish to neutralize their influence. Finally,  the $\ell_1$-penalized estimators we use for $\beta_0$ and $\mu_0$ are not conventional. The estimator of $\beta_0$ relies on a non-standard loss function and, to our knowledge, the properties of $\hat\beta$ that we need are not available in the literature, cf., e.g., \cite{vandegeer} and the references therein. The estimator of $\mu_0$ is close to the usual Lasso except for the weights that depend on $\hat \beta$. In general, discrepancy in the weights can induce an extra bias. 
Thus, it is not granted that such an estimator achieves properties close to the Lasso. We show below that it holds true under our assumptions. Our proof techniques may be of interest for other problems of similar type.

\medskip
Let  $\eta=(\beta,\mu)$ denote the vector of two nuisance parameters and recall that $Z = (D,X,Y)$. In what follows, we write for brevity $g(Z,\theta,\eta)$ instead of $g(Z,\theta,(\beta,\mu))$. In particular, for the value $\eta_0:=(\beta_0,\mu_0)$ we have 
\begin{equation}\label{main}
\mathbb{E}g(Z,\theta_0,\eta_0) = 0.
  \end{equation}
Hereafter, the notation $a  \lesssim b$ means that $a \leq cb$ for some constant $c>0$ independent of the sample size $n$. We denote by $\Phi$ and $\Phi^{-1}$ the cumulative distribution function and the quantile function of a standard normal random variable, respectively. We use the symbol $\mathbb{E}_n(\cdot)$ to denote the empirical average, that is  $\mathbb{E}_n(a) = n^{-1} \sum_{i=1}^n a_i$ for $a=(a_1,\dots,a_n)$. Finally, for a vector $\delta=(\delta_1,\dots,\delta_p) \in \mathbb{R}^p$ and a subset $S \subseteq \{1,\dots,p\}$ we consider the restricted vector
$\delta_S=(\delta_j\mathbb{I}(j\in S))_{j=1}^p$, where $\mathbb{I}(\cdot)$ denotes the indicator function, and we set $\norm{\delta}_0 := \text{Card}\left\{1\le j \le p: \delta_j \neq 0 \right\}$, $\norm{\delta}_1 := \sum_{j=1}^p \vert \delta_j \vert$, $\norm{\delta}_2 := \sqrt{ \sum_{j=1}^p \delta_j^2}$ and $\norm{\delta}_\infty := \underset{j=1,...,p}{\max} \vert \delta_j \vert$. 

\bigskip
We now state the assumptions used to prove the asymptotic results.

\begin{assumption}[Sparsity  restrictions]
\label{ass:dimres}
The nuisance parameter is sparse in the following sense:
\begin{align*}
\lVert \beta_0 \rVert_0 \leq s_\beta \text{,   } \lVert \mu_0 \rVert_0 \leq s_\mu
\end{align*}
for some integers $s_\beta, s_\mu\in [1,p]$.
\end{assumption}
\begin{assumption}[Conditions on function $h$] \label{assump:h}
Function $h$ is increasing, twice continuously differentiable on $\mathbb{R}$ and 
\begin{itemize}
\item[(i)] the second derivative $h''$ is Lipschitz on  any compact subset of $\mathbb{R}$,
\item[(ii)]  either  $\inf_{u\in \mathbb{R}}h'(u)\ge c_2 $  or $\norm{\beta_0}_1\le c_3$ where $c_2>0$ and $c_3>0$ are constants independent of $n$.
\end{itemize}
\end{assumption}
We also need some conditions on the distribution of data.  The random vectors $Z_i=(D_i,X_i,Y_i)$ are assumed to be i.i.d. copies of $Z=(D,X,Y)$  with $D\in \{0,1\}$, $X\in \mathbb{R}^p$ and $Y\in \mathbb{R}$. Throughout the paper, we assume that $Z$ depends on $n$, so that in fact we deal with a triangular array of random vectors. This dependence on $n$ is needed for rigorously stating asymptotical results since we consider the setting where the dimension $p=p(n)$ is a function of~$n$.  Thus, in what follows $Z$ is indexed by $n$ but  for brevity we typically suppress this dependence in the notation. On the other hand, all constants denoted by $c$ (with various indices) and $K$ appearing below 
are independent of $n$.

\begin{assumption}[Conditions on the distribution of data] \label{CovBound}
The random vectors $Z_i$ are i.i.d. copies of $Z=(D,X,Y)$  with $D\in \{0,1\}$, $X\in \mathbb{R}^p$ and $Y\in \mathbb{R}$ satisfying \eqref{eq:calibr-our}, \eqref{eq:mu0_OLS0}, \eqref{main} and the following conditions:\\
(i) There exist constants $K>0$ and $c_1>0$ such that
\begin{align*}
&\max \{\norm{X}_\infty,  |X^T\beta_0|,  |Y-X^T\mu_0| \}\le K \text{(a.s.)}, \quad \text{and} \quad  0<P(D=1) < 1.
\end{align*}
(ii) Non-degeneracy conditions. There exists $c_1>0$ such that for all $j=1,\dots, p$,
\begin{align*}
\min \bigg\{ &  \mathbb{E}\left( (Y - X^T \mu_0 -\theta_0)^2 \vert D = 1\right),  \ \mathbb{E}\left( h^2(X^T \beta_0) (Y - X^T \mu_0)^2 \vert D = 0 \right), \\
& \mathbb{E}\left((X^Te_j)^2 \vert D = 1\right), \ \mathbb{E}\left( h^2(X^T \beta_0)(X^Te_j)^2 \vert D = 0 \right), \\
& \mathbb{E}\left( (h'(X^T \beta_0))^2 (Y - X^T \mu_0)^2  (X^Te_j)^2 \vert D = 0 \right)  \bigg\} \ge c_1,
\end{align*}
where $e_j$ denotes the $j$th canonical basis vector in $\mathbb{R}^p$.
\end{assumption}

\begin{assumption}[Condition on the population Gram matrix of the control group] 
\label{ass:gram}
The population Gram matrix of the control group  
$$\Sigma := \mathbb{E}((1-D)X X^T)$$ 
is such that 
\begin{equation}\label{eq:gram}
 \underset{v\in \mathbb{R}^p\setminus \{0\}}{\min} \frac{v^T \Sigma v}{\norm{v}_2^2}  \ge \kappa_{\Sigma},
\end{equation}
where the minimal eigenvalue $\kappa_{\Sigma}$ is a positive number.
\end{assumption}
Note that in view of Assumption \ref{CovBound}, $\kappa_{\Sigma}$ is uniformly bounded: $\kappa_{\Sigma}\le e_1^T \Sigma e_1\le K^2$. 

\begin{remark}
	If (i) $c_p:=\sup_x P[D=1|X=x]<1$ and (ii) $\E[XX^T]$ is nonsingular, with smallest eigenvalue equal to $\lambda_{\min}$, then Assumption \ref{ass:gram} holds with $\kappa_\Sigma=(1-c_p)\lambda_{\min}$. Condition (i) is a slight reinforcement of Assumption \ref{ass:nonmadattreat}, which holds for instance (since $X$ is bounded) if $x\mapsto P[D=1|X=x]$ is continuous.
	\label{rem:prob_D} 
\end{remark}

\begin{assumption}[Dimension restrictions]
\label{ass:grow}
The integers $p, s=\max(s_\beta, s_\mu)\in [1,p/2]$ and the value $\kappa_{\Sigma}>0$ are functions of $n$ satisfying the following growth conditions:
\begin{itemize}
\item[(i)] $\displaystyle{\frac{s^2\log(p) }{ \kappa^2_{\Sigma}\sqrt{n} }}  \rightarrow 0$ as $n\to \infty$,
\item[(ii)] $s/\kappa_{\Sigma}= o(p)$ as $n\to \infty$,
\item[(iii)] $\log(p)= o(n^{1/3})$ as $n\to \infty$.   
\end{itemize}
\end{assumption}

Finally, we define the penalty loadings for estimation of nuisance parameters. The gradients of the estimating function with respect to the nuisance parameters are
\begin{align*}
\nabla_{\mu}  g(Z,\theta,\eta) &=    -\left[ D - (1-D) h(X^T\beta) \right] X, \\
\nabla_{\beta} g(Z,\theta,\eta) &= -(1-D) h'(X^T\beta) \left[Y- X^T \mu\right]X.
\end{align*}
For each $i=1,\dots,n$, we define the random vector $U_i \in \mathbb{R}^{2p}$ with entries corresponding to these gradients:
\[
U_{i,j} := \left\{
  \begin{tabular}{ll}
  $-\left[ D_i - (1-D_i) h(X_i^T\beta_0) \right] X_{i,j}$ & if $1 \le j \le p$, \\
  $-(1-D_i) h'(X_i^T\beta_0) \left[Y_i- X_i^T \mu_0\right]X_{i,j}
  $ & if $p+1 \le j \le 2p$,
  \end{tabular}
\right.
\]
where $X_{i,j}$ is the $j$th entry of $X_i$.

\begin{assumption}[Penalty Loadings] \label{ass:penloadings}
	Let $c > 1$ and $\gamma\in(0, 2p)$ be such that $\log(1/\gamma) \lesssim \log(p)$ and $\gamma=o(1)$ as $n\to\infty$. 
	%
	The penalty loadings for estimation of $\beta_0$ satisfy
	\begin{align*}
	\lambda &:= c \Phi^{-1}(1 - \gamma/2p) / \sqrt{n},\\
	 \psi_{j, \max} \ge \psi_{j} & \ge  \sqrt{n^{-1}\sum_{i=1}^n U_{i,j}^2}
	\quad \text{for   } j=1,\dots,p.
	\end{align*}
	The penalty loadings for estimation of $\mu_0$ satisfy
	\begin{align*}
	\lambda' &:= 2c \Phi^{-1}(1 - \gamma/2p) / \sqrt{n},\\
	\psi'_{j,\max} \ge \psi'_{j} & \ge  \sqrt{n^{-1}\sum_{i=1}^n U_{i,j+p}^2} \quad 
	\text{for   } j=1,\dots,p.
	\end{align*}
\end{assumption}

Here, the upper bounds on the loadings are	
	$$\psi_{j, \max}=\sqrt{\frac{1}{n} \sum_{i=1}^n \max_{|u|\le K}\left[(1-D_i) h(u) - D_i\right]^2 X_{i,j}^2} ,$$
$$
\psi'_{j,\max} =  \sqrt{\frac{1}{n} \sum_{i=1}^n (1-D_i)\max_{|u|\le K} \big(|h'(u)|^2 \left[Y_i - u\right]^2\big) X_{i,j}^2}
$$
representing feasible majorants for $\sqrt{n^{-1}\sum_{i=1}^n U_{i,j}^2}$ under our assumptions.


The values  $\sqrt{n^{-1}\sum_{i=1}^n U_{i,j}^2}$ depend on the unknown parameters. Thus, we cannot choose $\psi_{j}, \psi'_{j}$ equal to these values but we can take them equal to the upper bounds $\psi_{j}=\psi_{j, \max}$ and $\psi'_{j}=\psi'_{j, \max}$. A more flexible iterative approach of choosing feasible loadings is discussed in Appendix \ref{alg}.\\

The following theorem constitutes the main asymptotic result of the paper. 
\begin{theorem}[Asymptotic Normality of the Immunized Estimator] \label{AsyNormality}
Let Assumptions \ref{ass:dimres} - \ref{ass:penloadings} hold. Then the immunized estimator $\hat \theta$ defined in Equation \eqref{eq:thetaimmun} satisfies
\begin{equation*}
   \hat \sigma^{-1} \sqrt{n}( \hat \theta - \theta_0)  \overset{\mathcal{D}}{\to} \mathcal{N}(0, 1)\text{ as } n \rightarrow \infty,
\end{equation*}
where $\hat \sigma^2 := \frac{1}{n}\sum_{i=1}^n  g^2(Z_i, \hat \theta, \hat \eta) \Big(\frac{1}{n}\sum_{i=1}^n D_i\Big)^{-2}$ is a consistent estimator of the asymptotic variance, and $ \overset{\mathcal{D}}{\to}$ denotes convergence in distribution.
\end{theorem}
The proof is given in Appendix \ref{sub:proof_thm_as_nor}. An important point underlying the root-$n$ convergence and asymptotic normality of $\hat\theta$ is the fact that the expected gradient of $g$ with respect to $\eta$ is zero at $\eta_0$. In granting this property, we follow the general methodology of estimation in the presence of high-dimensional nuisance parameters  developed in \cite{BelloniChenChernozhukovHansen2012, BCH2012Treat, BelloniChernozhukovFenanderValHansen2014, ChernozhukovHansenSpindler2015b} among other papers by the same authors. 
The second important ingredient of the proof is to ensure that the estimator $\hat\eta$ converges fast enough to the nuisance parameter $\eta_0$. Its rate of convergence is given in the following theorem. 

\begin{theorem}[Nuisance Parameter Estimation] \label{the:nuisparam}
Under Assumptions \ref{ass:dimres}-\ref{ass:penloadings}, we have, with probability tending to 1 as $n\to\infty$,
\begin{eqnarray}
\lVert \hat \beta - \beta_0 \rVert_1 \lesssim \frac{s_\beta}{\kappa_\Sigma} \sqrt{ \frac{\log(p)}{ n} }, \label{ineq:norme1} \\
\lVert \hat \mu - \mu_0 \rVert_1 \lesssim \frac{s}{\kappa_\Sigma} \sqrt{ \frac{\log(p)}{ n} }. \label{ineq:norme1mu}
\end{eqnarray}
\end{theorem}

The proof is given in Appendix \ref{app:proof_nuis}. Theorem \ref{the:nuisparam}, together with the fact that $\hat\theta$ is immunized, implies that $\hat\beta$ and $\hat\mu$ have no asymptotic effect on $\hat\theta$. In related work, similar conditions to \eqref{ineq:norme1} and \eqref{ineq:norme1mu} have been imposed rather than established. We refer in particular to (36) of \cite{ChernozhukovHansenSpindler2015b} and Assumption 3-(ii) in \cite{Farrell2015}. Note, on the other hand, that contrary to \citeauthor{Farrell2015} (2015, see Assumption 3-(i)), we do not require that both $x\mapsto G(x'\hat\beta)$ and $x\mapsto x'\hat\mu$ are consistent for $x\mapsto P[D=1|X=x]$ and $x\mapsto \E[Y(0)|X=x]$. Theorem \ref{AsyNormality} only requires \eqref{main} to hold. By Theorem \ref{thm:double_rob_param} above, this is the case if either $P[D=1|X]=G(X^T\beta_0)$ or $\E[Y(0)|X]=X^T\mu_0$, in which case either $x\mapsto G(x'\hat\beta)$ or $x\mapsto x'\hat\mu$ is consistent, but not necessarily both.

\begin{remark}
The rate in \eqref{ineq:norme1mu} depends not only on the sparsity index $s_\mu$ of $\mu_0$, but on the maximum $s=\max(s_\beta, s_\mu)$. This is natural since one should account for the accuracy of the preliminary estimator $\hat\beta$ used to obtain $\hat\mu$. 
\label{rem:rate_beta_mu}
\end{remark}

\begin{remark}
	Inspection of the proof shows that Theorem \ref{the:nuisparam} remains valid under weaker assumptions, namely, we can replace Assumption~\ref{ass:grow} on $s,p,\kappa_\Sigma$ by the condition \eqref{ass:rud}. We also note that Assumption~\ref{ass:grow}(i) in Theorem~\ref{AsyNormality} can be modified to $(s/\sqrt{n})\log(p)\to 0$ if $\kappa_\Sigma$ is a constant independent of $n$ as it is assumed, for example, in \cite{Wager2019}. Such a modification would require a substantially more involved proof but only improves upon considering relatively non-sparse cases. This does not seem of much added value for using the methods in practice when the sparsity index is typically small. Moreover, in the high-dimensional scenario we find it more important  to specify the dependency of the growth conditions on $\kappa_\Sigma$. 
\label{rem:rate2}
\end{remark}

\begin{remark}
	The accuracy of $\hat\beta$, $\hat\mu$ and ultimately $\hat{\theta}$ may be affected by the exact value of $c_p:=\sup_x P[D=1|X=x]$. When $c_p$ is close to 1, our bounds on the performance of estimators deteriorate, see Theorem \ref{the:nuisparam} and Remark \ref{rem:prob_D}. Such a phenomenon could be expected. If $P[D=1|X=x]$ gets closer to 1 it becomes more difficult to estimate the distribution of $Y$ conditional on $D=0$ and $X$, which is required  to estimate $\E[Y(0)|D=1]$ and, in turn, $\theta_0$. Related to this, \cite{khan2010irregular} show that, in the absence of further restrictions on $x\mapsto P[D=1|X=x]$ and $x\mapsto E(Y(0)|X=x)$, the condition $\sup_x P[D=1|X=x]<1$ is necessary for root-$n$ consistency in the problem of estimation of the ATT \citep[see Theorem 4.1 in][]{khan2010irregular}.\footnote{Their result is on the average treatment effect (ATE), rather than on the ATT, but their proof can  be simply adapted to the ATT.} 
	\label{rem:prob_D2}
\end{remark}

\section{Monte Carlo Simulations}
\label{sec:MCXP}

\begin{figure}[H]
\centering
\includegraphics[width=1 \textwidth]{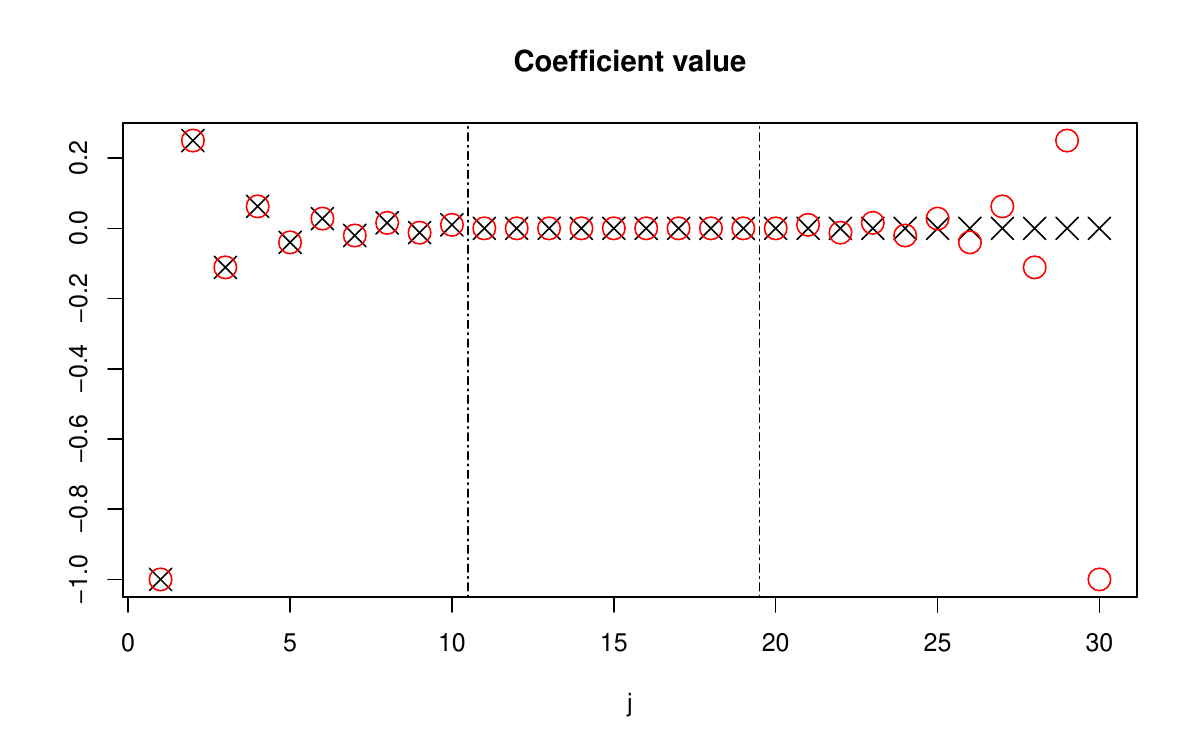}
\begin{minipage}{0.95 \textwidth}
{\footnotesize
Notes: In this example, $\rho_\gamma=\rho_\mu=1$. The central region of the graph represents the coefficients $\gamma_0$ and $\mu_0$ associated with variables that do not play a role either in the selection equation or in the outcome equation. The left region shows the coefficients associated with variables that are important for both equations. In the right region, only entries in $\mu_0$ is different from 0, meaning that the variables determine the outcome equation but not the selection equation.}
\end{minipage}
\caption{\label{SparsityPattern} Sparsity patterns of $\gamma_0$ (crosses) and $\mu_0$ (circles). }
\end{figure}

The aim of this experiment is two-fold: illustrate the better properties of the immunized estimator over the naive plug-in, and compare it with other estimators. In particular, we compare it with a similar estimator proposed by \cite{Farrell2015}. 
We consider the following DGP. The $p$ covariates are distributed as $X \sim \mathcal{N}(0,\Sigma)$, where the 
$(i,j)$th element of the covariance matrix satisfies $\Sigma_{i,j}=0.5^{\vert i-j \vert}$. The treatment equation follows a logit model, $P(D=1|X) =\Lambda\left( X^T \gamma_0 \right)$ with $\Lambda(u) = 1/(1+\exp(-u))$. The potential outcomes satisfy
$$Y(0) = \exp(X^T\mu_0) + \varepsilon, \quad Y(1) = Y(0) + \zeta_0 X^T\gamma_0,$$  
where $\zeta_0$ is a constant, $\varepsilon \sim \mathcal{N}(0,1)$ and $\varepsilon$ is independent of $(D,X)$. We assume the following form for the $j$th entry of $\gamma_0$ and $\mu_0$:
\begin{equation*}
\gamma_{0j} = \left\{
      \begin{aligned}
& \rho_\gamma (-1)^j/ j^2 \text{ \; if  } j \le 10 \\
& 0 \text{\hspace{2cm} otherwise}
\end{aligned}
\right.  \text{,  }\; \mu_{0j} = \left\{
      \begin{aligned}
& \rho_\mu (-1)^{j}/ j^2 \text{ \hspace{3cm} if   } j \le 10 \\
& \rho_\mu (-1)^{j+1} / (p-j+1)^2 \text{ \hspace{.9cm} if   } j \ge p-9 \\
& 0 \text{ \hspace{4.8cm} otherwise,}
\end{aligned}
\right. 
\end{equation*}
where $\rho_\gamma$ and $\rho_\mu$ are positive constants.
We are thus in an strictly sparse setting for both equations where only ten covariates play a role in the treatment assignment and twenty in the outcome. Figure \ref{SparsityPattern} depicts the precise pattern of the corresponding coefficients for $p=30$. The constants $\rho_\gamma$ and $\rho_\mu$  fix the signal-to-noise ratio. More precisely, $\rho_\gamma$ is set so that $R^2=0.3$ in the latent model for $D$, and $\rho_\mu$ is set so that $R^2=0.8$ in the model for $Y(0)$. Finally, we let $\zeta_0= [V\left( \exp(X^T\mu_0) \right)/5V\left(Y(0)\right)]^{1/2}$. This impies that the variance of the individual treatment effect $\zeta_0 X'\gamma_0$ is one fifth of the variance of $Y(0)$. In this set-up, the ATT satisfies $E[Y(1) -Y(0) \vert D=1] = \zeta_0E[Z\Lambda(Z)]/E[\Lambda(Z)]$, with $Z \sim \mathcal{N}(0,\gamma_0^T \Sigma \gamma_0)$. We compute it using Monte Carlo simulations.

We consider several estimators of the ATT. The first is the naive plug-in estimator defined in  \eqref{eq:naiveplugin}. Next, we consider our proposed estimator defined in \eqref{eq:thetaimmun}, with $H(x)=h(x)=\exp(x)$. We also consider the estimator proposed by \cite{Farrell2015}. This estimator is also defined by \eqref{eq:thetaimmun}, but $\hat{\beta}$ and $\hat{\mu}$ therein are  obtained by a Logit Lasso and an unweighted Lasso regression, respectively. For these three estimators, the penalty loadings for the first-step estimators are set as in Appendix \ref{alg}. The last estimator, called the oracle hereafter, is our low-dimensional estimator defined by \eqref{eq:theta_ld}, where in the first step we only include the ten covariates affecting the treatment. For all estimators, we construct 95\% confidence estimator on the ATT using the normal approximation and asymptotic variance estimators. We estimate the asymptotic variance of the naive plug-in estimator making as if we were in a low-dimensional setting. This means that this estimator would have an asymptotic coverage of 0.95 if $p$ remained fixed.

\medskip
In our DGP, the variables $X_j$ with $j\geq p-9$ matter in the outcome equation but are irrelevant in the treatment assignment rule. Given that the propensity score is correctly specified, the four estimators are consistent. The oracle estimator should be the most accurate since it incorporates the information on which covariates matter in the propensity score. Because the balancing program misses some covariates that are relevant for the outcome variable (namely, the $X_j$ with $j\geq p-9$),   the naive plug-in estimator is expected to be asymptotically biased. The immunized procedure should correct for this bias.

\begin{table}[H]
\caption{Monte-Carlo simulations}
\label{table:simulations}
\centering
\begin{small}
\begin{tabular}{lccccccccc}
\multicolumn{10}{c}{}\\
\toprule
& \multicolumn{3}{c}{$n=500$} &\multicolumn{3}{c}{$n=1,000$}  &  \multicolumn{3}{c}{$n=2,000$}  \\
& RMSE & Bias & CR & RMSE & Bias & CR & RMSE & Bias & CR   \\
\cmidrule(lr){2-4}\cmidrule(lr){5-7}\cmidrule(lr){8-10}
& \multicolumn{9}{c}{$p=50$} \\
\cmidrule(lr){2-4}\cmidrule(lr){5-7}\cmidrule(lr){8-10}
Naive plug-in  &  0.312  &  0.264  &  0.62  &  0.228  &  0.195  &  0.601  &  0.169  &  0.144  &  0.608  \\
Immunized  &  0.186  &  0.102  &  0.872  &  0.121  &  0.059  &  0.907  &  0.084  &  0.03  &  0.907  \\
Farrell  &  0.197  &  0.117  &  0.857  &  0.132  &  0.075  &  0.885  &  0.093  &  0.046  &  0.887  \\
Oracle  &  0.202  &  -0.017  &  0.929  &  0.143  &  -0.025  &  0.938  &  0.105  &  -0.03  &  0.935  \\
& \multicolumn{9}{c}{$p=200$} \\
\cmidrule(lr){2-4}\cmidrule(lr){5-7}\cmidrule(lr){8-10}
Naive plug-in  &  0.318  &  0.274  &  0.587  &  0.238  &  0.208  &  0.557  &  0.179  &  0.156  &  0.544  \\
Immunized  &  0.185  &  0.108  &  0.883  &  0.125  &  0.066  &  0.897  &  0.085  &  0.037  &  0.911  \\
Farrell  &  0.195  &  0.121  &  0.87  &  0.135  &  0.081  &  0.873  &  0.095  &  0.052  &  0.882  \\
Oracle  &  0.194  &  -0.023  &  0.936  &  0.141  &  -0.029  &  0.946  &  0.108  &  -0.034  &  0.934  \\
& \multicolumn{9}{c}{$p=500$} \\
\cmidrule(lr){2-4}\cmidrule(lr){5-7}\cmidrule(lr){8-10}
Naive plug-in  &  0.339  &  0.297  &  0.532  &  0.247  &  0.216  &  0.522  &  0.185  &  0.165  &  0.494  \\
Immunized  &  0.202  &  0.128  &  0.841  &  0.128  &  0.07  &  0.881  &  0.087  &  0.044  &  0.912  \\
Farrell  &  0.212  &  0.141  &  0.81  &  0.14  &  0.086  &  0.854  &  0.098  &  0.059  &  0.866  \\
Oracle  &  0.205  &  -0.019  &  0.927  &  0.146  &  -0.033  &  0.931  &  0.105  &  -0.032  &  0.938  \\
& \multicolumn{9}{c}{$p=1,000$} \\
\cmidrule(lr){2-4}\cmidrule(lr){5-7}\cmidrule(lr){8-10}
Naive plug-in  &  0.345  &  0.309  &  0.485   &  0.258  &  0.23  &  0.478   &  0.194  &  0.175  &  0.449   \\
Immunized  &  0.199  &  0.133  &  0.835   &  0.135  &  0.082  &  0.862   &  0.09  &  0.051  &  0.885   \\
Farrell  &  0.211  &  0.146  &  0.807   &  0.146  &  0.097  &  0.823   &  0.101  &  0.066  &  0.853   \\
Oracle  &  0.194  &  -0.017  &  0.947   &  0.14  &  -0.022  &  0.943   &  0.103  &  -0.026  &  0.939  \\
\bottomrule
\multicolumn{10}{p{420pt}}{{\footnotesize Notes: RMSE and CR stand respectively for root mean squared error and coverage rate. The nominal coverage rate is 0.95. The results are based on 10,000 simulations for each $(n,p)$. The naive plug-in and immunized estimators are defined in  \eqref{eq:naiveplugin} and \eqref{eq:thetaimmun}, respectively. ``Farrell'' is the estimator considered by \cite{Farrell2015}.``Oracle'' is defined by \eqref{eq:theta_ld}, where in the first step we only include the ten covariates affecting the treatment.}}
\end{tabular}
\end{small}
\end{table}

\medskip
Table \ref{table:simulations} displays the results. We consider several values of $n$ and $p$ that approximate a relatively high-dimensional setting. For every couple $(n,p)$, we report the root mean squared error (RMSE), the bias and the coverage rate of the confidence intervals associated to each estimator. Our estimator performs well in all settings, with a correct coverage rate and often the lowest RMSE over all estimators. The oracle has always a coverage rate close to 0.95 and a bias very close to 0, as one could expect, but it does not always exhibit the lowest RMSE. This is because, intuitively, the immunized estimator and that of \cite{Farrell2015} trade off variance with some bias in their first steps, sometimes resulting in slightly lower RMSE on the final estimator. Note that the bias of the final estimator, though asymptotically negligible, results in a slight undercoverage of the confidence intervals. Yet, even with $n=500$ and $p=1,000$, the coverage rate is still of 0.84, much higher than that of the naive plug-in estimator (0.485). This estimator exhibits a large bias and a low coverage rate even for $p=50$ and $n=2,000$. This shows the importance of correcting for the bias of the first-step estimator, even when $p/n$ is quite small.  Finally, the estimator of \cite{Farrell2015} exhibits similar performance as the immunized estimator, though it displays a slightly larger RMSE and smaller coverage rate with this particular DGP.

\section{Empirical Applications}
\label{sec:applications}

\subsection{Job Training Program}
\label{sub:NSW}

We revisit \cite{lal86}, who examines the ability of econometric methods to recover the causal effect of employment programs.\footnote{For more discussion on the NSW program and the controversy regarding econometric estimates of  causal effects based on nonexperimental data, see  \cite{lal86} and the subsequent contributions by  \cite{dw99,DehejiaWahba2002,st05}.} This dataset was first built to assess the impact of the National Supported Work (NSW) program. The NSW is a transitional, subsidized work experience program targeted towards people with longstanding employment problems: ex-offenders, former drug addicts, women who were long-term recipients of welfare benefits and school dropouts. The quantity of interest is the average effect of the program for the participants on 1978 yearly earnings. The treated group gathers people who were randomly assigned to this program from the population at risk (with a sample size of $n_1 =185$). Two control groups are available. The first one comes from the Panel Study of Income Dynamics (PSID) (sample size $n_0 =2,490$). The second one comes from the experiment (sample size $n_0=260$) and  is therefore directly comparable to the treated group. It provides us with a benchmark for the ATT. Hereafter, we use the group of participants and the PSID sample to compute our estimator and compare it with other competitors and the experimental benchmark.

\medskip
To allow for a flexible specification, we follow \cite{Farrell2015} by taking the raw covariates of the dataset (age, education, black, hispanic, married, dummy variable of no degree, income in 1974, income in 1975, dummy variables of  no earnings in 1974 and in 1975), two-by-two-interactions between the continuous and dummy variables, two-by-two interactions between the dummy variables and powers up to degree 5 of the continuous variables. Continuous variables are linearly rescaled to $[0,1]$. We end up with 172 variables to select from. The experimental benchmark for the ATT estimate is \$1,794, with a standard error of 671. We compare several estimators: the naive plug-in estimator, the immunized plug-in estimator, the doubly-robust estimator of \cite{Farrell2015}, the double-post-selection linear estimator of \cite{BCH2012Treat}, and the plain OLS estimator including all the covariates. For the four penalized estimators, the penalty loadings on the first-step estimators are set as in Appendix \ref{alg}.

\medskip
Table \ref{ATT_Results} displays the results. Our immunized estimator and that of \cite{Farrell2015} give credible values for the ATT with respect to the experimental benchmark, with also similar standard errors.\footnote{\cite{Farrell2015}'s estimate shown in Table \ref{ATT_Results} differs from that displayed in Farrell's paper because contrary to him, we have not automatically included education, the dummy of no degree and the 1974 income in the set of theory pre-selected covariates. When doing so, the results are slightly better but not qualitatively different for this estimator. We chose not to do so as it would bias the comparison with other estimators, which do not include a set of pre-selected variables. } Notably, only these two estimators out of the five considered display a significant positive impact as the experimental benchmark. The immunized estimator estimator offers a substantial improvement on bias and standard error over the naive plug-in estimator, in line with the evidence from the Monte Carlo experiment. Finally, the OLS estimator in Column (6) is a benchmark of a very simple procedure that does not use any selection at all. As no surprise, this estimator has a substantially larger standard error than the other considered estimators.

\begin{table}[H]
\begin{center}
\begin{small}
\caption{ATT estimates on the NSW program.} \label{ATT_Results}
\begin{tabular}{@{\extracolsep{5pt}}lccc} 
\multicolumn{4}{c}{}\\
\toprule
 & \multicolumn{3}{c}{Estimator} \\ 
\cline{2-4} 
\\[-1.8ex] & Experimental & Naive &Immunized \\ 
 & benchmark & plug-in & estimator \\ 
 & (1) & (2) & (3) \\ 
\cline{2-4} \\[-1.8ex]
Point estimate & 1,794.34 & 401.89 & 1,608.99   \\ 
Standard error  & (671.00) & (746.07) &  (705.38)  \\ 
95\% confidence interval & [519;\,3,046] & [-1,060;\,1,864] & [226;\,2991]  \\
$\#$ variables in propensity score & none &  9 & 9 \\ 
$\#$ variables in outcome equation & none & none &  12  \\
\hline \\ 
& Farrell & BCH & OLS \\ 
 &  (2015) & (2014) & estimator \\ 
 & (4) & (5) & (6) \\ 
\cline{2-4} \\[-1.8ex]
Point estimate & 1420.43 &  226.82 & 83.17 \\ 
Standard error  & (670.32)  & (867.51) & (1,184.48)  \\ 
95\% confidence interval & [107;\,2734] & [-1473;\,1927] & [-2,238;\,2,405] \\
$\#$ variables in propensity score    &  3 & 16 & none \\ 
$\#$ variables in outcome equation    & 13 & 10 & 172 \\
\bottomrule
\multicolumn{4}{p{400pt}}{{\footnotesize Notes: For details on the estimators, see the text. Standard errors and confidence intervals are based on the asymptotic distribution.}}
\end{tabular} 
\end{small}
\end{center}
\end{table}

\subsection{California Tobacco Control Program}
\label{sub:tobacco}

Proposition 99 is one of the first and most ambitious large-scale tobacco control program, implemented in 1989 in California. It includes a vast array of measures, including an increase in cigarette taxation of 25 cents per pack, and a significant effort in prevention and education. In particular, the tax revenues generated by Proposition 99 were used to fund anti-smoking campaigns. \cite{AbadieDiamondHainmueller2010} analyzes the impact of the law on tobacco consumption in California. Since this program was only enforced in California it is a nice example where the synthetic control method applies whereas more standard public policy evaluation tools cannot be used. It is possible to reproduce a synthetic California by reweighting other states so as to mimic California's behavior.

\begin{figure}[H]
\centering
\includegraphics[width=.95\textwidth]{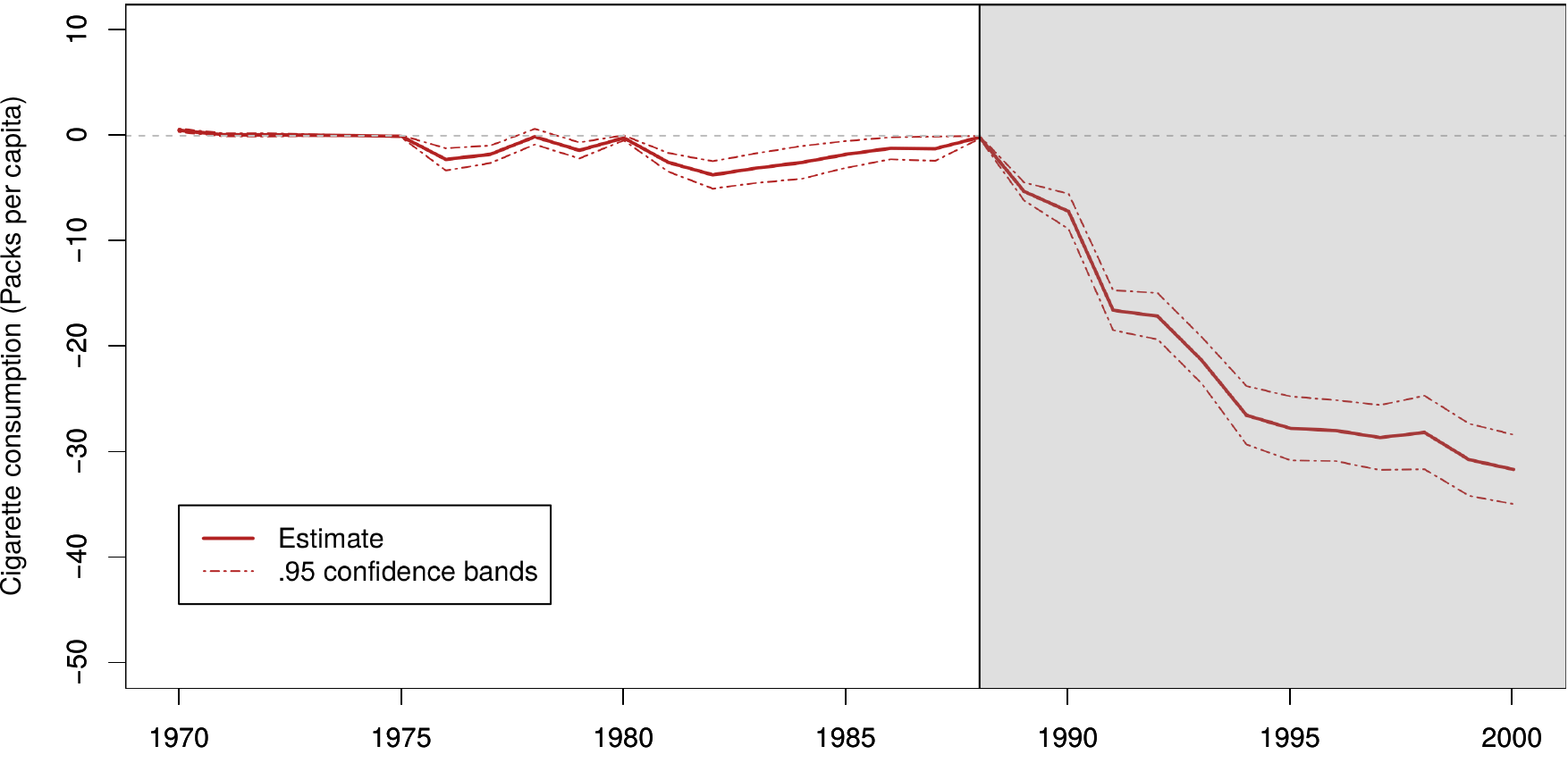}
\begin{minipage}{0.95\textwidth}
\medskip
	{\footnotesize Notes: the shaded area represents the post-treatment period. The 95\% confidence interval is based on the asymptotic approximation.}
\end{minipage}
\caption{The effect of Proposition 99 on per capita tobacco consumption.}
\label{fig:Proposition99TreatmentEffect} 
\end{figure}

\medskip
For this purpose, \cite{AbadieDiamondHainmueller2010} consider the following covariates: retail price of cigarettes, state log income per capita, percentage of the population between 15 and 24, per capita beer consumption (all 1980-1988 averages). Cigarette consumptions for  the years between 1970 and 1975, 1980 and 1988 are also included. Using the same variables, we conduct the same analysis with our estimator. Figure \ref{fig:Proposition99TreatmentEffect} displays the estimated effect of Proposition 99 using the immunized estimator.

We find almost no effect of the policy over the pre-treatment period, giving credibility to the counterfactual employed. A steady decline takes place after 1988, and in the long-run, tobacco consumption is estimated to have decreased by about 30 packs per capita per year in California as a consequence of the policy. The variance is larger towards the end of the period because covariates are measured in the pre-treatment period and they become less relevant as predictors. Note also that by construction, including 1970 to 1975, 1980 and 1988 cigarette consumptions among the covariates yields a very good fit at these dates because of the immunization procedure. The fit is not perfect, however, because of the shrinkage induced by the $\ell_1$-penalization.

\begin{figure}[H]
\centering
\includegraphics[scale=0.5]{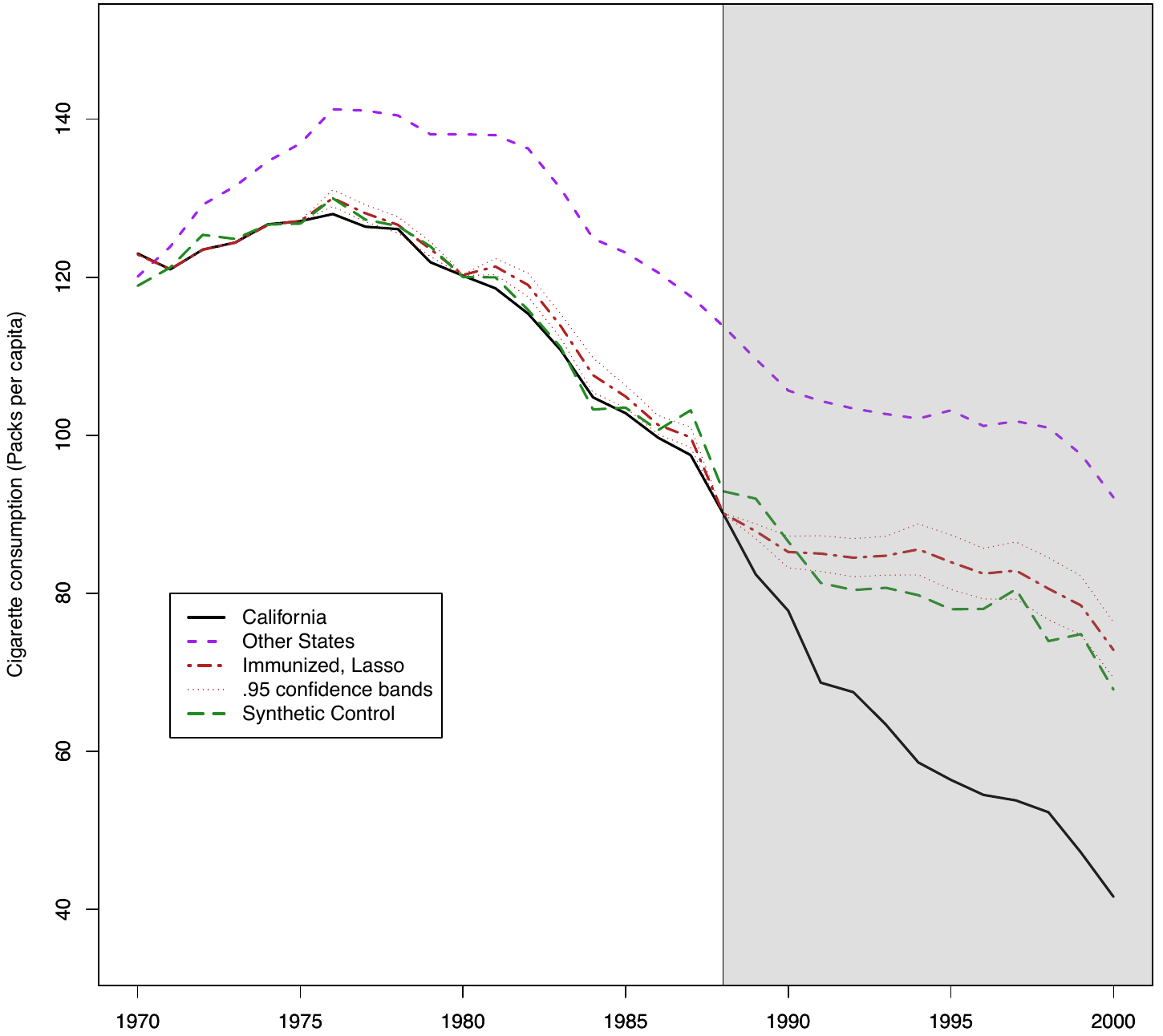}
\begin{minipage}{0.95\textwidth}
\medskip
	{\footnotesize Notes:  The solid black line is California tobacco consumption as in the data. The dotted purple line is a simple average of other U.S. states. The dashed red line is the immunized estimator as presented in the paper along with the 95\% confidence bands. The dashed green line is synthetic California.}
\end{minipage}
\caption{Cigarette consumption in California, actual and counterfactual}
\label{fig:CaliforniaImmunizedvADH} 
\end{figure}

\medskip
Figure \ref{fig:CaliforniaImmunizedvADH} displays a comparison between the immunized estimator and the synthetic control method. The dashed green line is the synthetic control counterfactual. It does not  match exactly the plot of \cite{AbadieDiamondHainmueller2010}, in which the weights given to each predictors are optimized to fit best the outcome over the whole pre-treatment period. Instead, the green curve in Figure \ref{fig:CaliforniaImmunizedvADH} optimizes the predictor weights using only the years 1970 through 1975, 1980 and 1988. This strategy brings a fairer comparison with our estimator that does not use California's per capita tobacco consumption outside those dates to optimize the fit. In such a case, the years from 1976 to 1987, excluding 1980, can be used as sort of placebo tests. 

Both our estimator and the synthetic control are credible counterfactuals, as they are able to closely match California pre-treatment tobacco consumption. They offer a sizable improvement over a sample average over the U.S. that did not implement any tobacco control program. Furthermore, even if our estimator gives a result relatively similar to the synthetic control, it displays a smoother pattern especially towards the end of the 1980s. The estimated treatment effect appears to be larger with the immunized estimate than with the synthetic control. However, it is hard to conclude that this difference is significant because in the absense of any asymptotic theory on the synthetic control estimator it is unclear how one could make a test on the difference between the two. 
In fact, the availability of standard asymptotic approximation for confidence intervals is to the advantage of our method.

\section{Conclusion} 
\label{sec:conclusion}

In this paper, we propose an estimator that makes a link between the synthetic control method, typically used with aggregated data and $n_0 $ smaller than or of the same order as $p$, and treatment effect methods used for micro data for which $n_0 \gg p$. Our method accommodates both settings. In the low-dimensional regime, it pins down one of the solutions of the synthetic control problem, which admits an infinity of solutions. In the high-dimensional regime, the estimator is a regularized and immunized version of the low-dimensional one and then differs from the synthetic control estimator. The simulations and applications suggest that it works well in practice. 

\medskip
In our study, we have focused on specific procedures based on $\ell_1$-penalization and proved that they achieve good asymptotic behavior in possibly high-dimensional regime under sparsity restrictions. Other types of estimators could be explored using these ideas. For example, in the high-dimensional regime, our strategy can be used with the whole spectrum  of sparsity-related penalization techniques, such as group Lasso, fused Lasso, adaptive Lasso, Slope, among other.


\bibliography{biblio}

\appendix


\section{Algorithm for Feasible Penalty Loadings}
\label{alg}

Consider the ideal penalty loadings for estimation of $\beta_0$ defined as
\begin{align*}
\lambda &:= c \Phi^{-1}(1 - \gamma/2p) / \sqrt{n}\\
\bar \psi_{j} &:=  \sqrt{\frac{1}{n} \sum_{i=1}^n \left[(1-D_i) h(X_i^T \beta_0) - D_i\right]^2 X_{i,j}^2} \text{   for   } j=1,...,p,
\end{align*}
and the ideal penalty loadings for estimation of $\mu_0$ :
\begin{align*}
\lambda' &:= 2c \Phi^{-1}(1 - \gamma/2p) / \sqrt{n}\\
\bar \psi'_{j} &:=  \sqrt{\frac{1}{n} \sum_{i=1}^n (1-D_i)h'(X_i^T \beta_0)^2 \left[Y_i - X_i^T \mu_0\right]^2 X_{i,j}^2} \text{  for   } j=1,...,p.
\end{align*}
Here $c > 1$ is an absolute constant, $\gamma>0$ is a tuning parameter while $\beta_0$ and $\mu_0$ are the true coefficients. We follow \cite{BCH2012Treat} and set $\gamma=.05$ and $c=1.1$.

\medskip
We first estimate the ideal penalty loadings $\left\{ \bar \psi_{j} \right\}^{p}_{j=1}$ of the balancing step using the following algorithm.
Set a small constant $\epsilon>0$ and a maximal number of iterations $k_0$.
\begin{enumerate}
\item Start by using a preliminary estimate $\beta^{(0)}$ of $\beta_0$. For example, take $\beta^{(0)}$ with the entry corresponding to the intercept equal to $\log(\sum_{i=1}^n D_i / \sum_{i=1}^n (1-D_i))$ and all other entries equal to zero. Then, for all $j=1,...,p$, set $$\tilde \psi_j^{(0)} = \sqrt{\frac{1}{n} \sum_{i=1}^n\left[ (1-D_i)h(X_i^T \beta^{(0)}) - D_i \right]^2 X_{i,j}^2}.$$
At step $k$, set $\tilde \psi_j^{(k)} = \sqrt{\frac{1}{n} \sum_{i=1}^n\left[ (1-D_i)h(X_i^T \beta^{(k)})-D_i\right]^2 X_{i,j}^2}$,  $j=1,...,p$.
\item Estimate the model by the penalized balancing Equation \eqref{betaLasso} using the penalty level $\lambda $ and penalty loadings found previously, to obtain $\hat \beta^{(k)}$.
\item Stop if $\underset{j=1,...,p}{\max} \vert \tilde \psi_j^{(k)}-\tilde \psi_j^{(k-1)} \vert \leq \epsilon$  or if $k > k_0$. Set $k=k+1$ and go to step 1 otherwise.
\end{enumerate}

Asymptotic validity of this approach is established analogously to \cite[Lemma 11]{BelloniChenChernozhukovHansen2012}. Estimation of the penalty loadings $\bar\psi'_{j}$ on the immunization step follows a similar procedure where we replace $\beta_0$ in the formula for $\bar\psi'_{j}$ by its estimator obtained on the balancing step.

\section{Proofs}

\subsection{Proof of Theorem \ref{thm:double_rob0}} 
\label{sub:theorem_ref_thm_double_rob}

First, note that we have $\mathbb{E} \left[(1-D)w(X)X\right] = \mathbb{E} \left[ D X\right]$. As a result, for any $\mu \in \mathbb{R}^p$, 
$$\mathbb{E}\left[ \left(D - (1-D)w(X) \right) (Y-X^T\mu)\right] = \mathbb{E}\left[ \left(D - (1-D)w(X) \right) Y\right].$$
Since $(1-D)Y = (1-D)Y(0)$ and $DY = D Y(1)$, the value $\theta_0$ satisfies the moment condition  \eqref{ATTEstimMoment} if and only if
\begin{equation*}
\mathbb{E}\left[ w(X) Y (1-D) \right] = \mathbb{E}\left[  D Y(0) \right].
\end{equation*}
By the Mean Independence assumption, $\mathbb{E}(D\vert X) \mathbb{E}(Y(0)\vert X)=\mathbb{E}(DY(0)|X)$. Thus,
\begin{equation}\label{mi}
\mathbb{E}\left[w(X) Y (1-D)\right] = \mathbb{E}\left[\mathbb{E}(w(X) Y(0) (1-D)|X)\right] = \mathbb{E}\left[w(X) (1-\mathbb{E}(D\vert X) )\mathbb{E}(Y(0)\vert X) \right].
\end{equation}
We consider the two cases of the theorem separately.
\begin{enumerate}
\item In the linear case $\mathbb{E} (Y(0) \vert X) = X^T \mu_0$ we have 
\begin{align*}
\mathbb{E}\left[w(X) Y (1-D)\right] &= \mathbb{E}\left[w(X) (1-D) X^T \mu_0 \right]\\
&= \mathbb{E}\left[DX^T \mu_0 \right]\\
&= \mathbb{E}\left[D \mathbb{E}(Y(0) \vert X) \right]\\
&= \mathbb{E}\left[DY(0) \right].
\end{align*}
The first equality here is due to \eqref{mi}. The second equality follows from the fact that  $\mathbb{E}[(1-D)w(X)X] = \mathbb{E} [DX]$. The last equality uses the Mean Independence assumption.

\item Propensity score satisfies $P(D=1|X)=w(X)/(1+w(X))$. In this case, using \eqref{mi} we have
\begin{align*}
\mathbb{E}\left[w(X) Y (1-D)\right] 
&= \mathbb{E}\left[w(X)(1-P(D=1|X)) \mathbb{E}(Y(0)\vert X) \right]\\
&= \mathbb{E}\left[P(D=1|X) \mathbb{E}(Y(0)\vert X) \right]\\
&= \mathbb{E}\left[\mathbb{E}(D\vert X) \mathbb{E}(Y(0)\vert X) \right]\\
&= \mathbb{E}\left[DY(0) \right],
\end{align*}
where the last equality follows from the Mean Independence assumption.
\end{enumerate}
\hfill$\square$


\subsection{Proof of Theorem \ref{AsyNormality}}
\label{sub:proof_thm_as_nor}

Denote the observed data by $Z_i = (Y_i,D_i,X_i)$, and by $\pi_0$ the probability of being treated: $\pi_0 := P(D=1)$.  The estimating moment function for $\theta_0$ is $g(Z,\theta,\eta):= [D - (1-D) h(X^T \beta) ] [Y - X^T\mu] - D \theta$. Recall that we define $(\theta_0, \eta_0)$ as the values satisfying:
\begin{equation*}
\mathbb{E}g(Z,\theta_0,\eta_0) = 0.
\end{equation*}
All these quantities depend on the sample size $n$ but for the sake of brevity we suppress this dependency in the notation except for the cases when we need it explicitly.

\medskip
By the Taylor expansion and the linearity of the estimating function $g$ in $\theta$, there exists $t \in (0,1)$ such that
\begin{align*}
\mathbb{E}_n[g(Z,\hat \theta,\hat \eta)] &= \E_n\left[g(Z,\theta_0,\hat \eta)\right] + \hat \pi (\theta_0 - \hat \theta)\\
&= \hat \pi (\theta_0 - \hat \theta) + \E_n\left[g(Z,\theta_0, \eta_0)\right] + (\hat \eta - \eta_0)^T \E_n\left[\nabla_{\eta} g(Z,\theta_0, \eta_0)\right]\\
&+ \frac{1}{2} (\hat \eta - \eta_0)^T  \mathbb{E}_n\left[\nabla^2_{\eta} g(Z,\theta_0, \tilde \eta)\right] (\hat \eta - \eta_0),
\end{align*}
where $\tilde \eta := t \eta_0 + (1-t) \hat \eta$ and $\hat \pi  =\frac{1}{n}\sum_{i=1}^{n}D_i$. The immunized estimator satisfies $\E_n[g(Z,\hat \theta,\hat \eta)] = 0$. Thus, we obtain
\begin{align*}
\hat \pi \sqrt{n} (\hat \theta - \theta_0) &= \underbrace{\sqrt{n}  \mathbb{E}_n\left[g(Z,\theta_0, \eta_0)\right]}_{:=I_1} + \underbrace{\sqrt{n}(\hat \eta - \eta_0)^T \mathbb{E}_n\left[\nabla_{\eta} g(Z,\theta_0, \eta_0)\right]}_{:= I_2}\\
&+ \underbrace{2^{-1}\sqrt{n}  (\hat \eta - \eta_0)^T  \mathbb{E}_n\left[\nabla^2_{\eta}  g(Z,\theta_0, \tilde \eta) \right](\hat \eta - \eta_0)}_{:=I_3}.
\end{align*}
 Now, to prove Theorem \ref{AsyNormality} we proceed as follows. First, we show that $I_1$ converges in distribution to a zero mean normal random variable with variance $\mathbb{E}g^2(Z,\theta_0, \eta_0)$ (Step 1 below), while $I_2$ and $I_3$ tend to zero in probability (Steps 2 and 3). This and the fact that $\hat \pi \to \pi_0$ (a.s.) imply that $\sqrt{n} (\hat \theta - \theta_0) $ is asymptotically normal with some positive variance, which in turn implies that $\hat \theta \to \theta_0$ in probability. Finally, using this property we prove that $\E_n\left[g^2(Z_i, \hat \theta, \hat \eta)\right] \to \mathbb{E}g^2(Z,\theta_0, \eta_0)$ in probability (Step 4). Combining Steps 1 to 4 and using Slutsky lemma leads to the result of the theorem. 
Thus, to complete the proof of the theorem it remains to establish Steps 1 to 4.

\subsubsection*{Step 1} 

In this part, we write $g_{i,n} := g(Z_i,\theta_0,\eta_0)$ (making the dependence on $n$ explicit in the notation). Recall that $\mathbb{E}g_{i,n} =0$. We apply the Lindeberg-Feller central limit theorem for triangular arrays by checking a Lyapunov condition. Since $(g_{i,n})_{1\leq i\leq n}$ are i.i.d., it suffices to prove that
\begin{equation}
    \limsup_{n\to \infty} \frac{\mathbb{E}(g_{1,n}^{2+\delta})}{(\mathbb{E}g_{1,n}^2)^{1+\delta/2}}<\infty
\label{eq:Lyapunov}
\end{equation}
for some $\delta>0$.
Assumption \ref{CovBound}(i) implies that $\mathbb{E}(g_{1,n}^{2+\delta})$ is bounded uniformly in $n$. Moreover,
\begin{align*}
\mathbb{E}(g_{1,n}^2) = \pi_0 \mathbb{E}\left( (Y(1) - X^T \mu_0 -\theta_0)^2 \vert D = 1 \right) +(1-\pi_0) \mathbb{E}\left( h(X^T \beta_0)^2 (Y(0) - X^T \mu_0)^2 \vert D = 0 \right).
\end{align*}
Due to Assumption \ref{CovBound}(ii) we have $\mathbb{E}(g_{1,n}^2)\ge c_1$, where $c_1>0$ does not depend on $n$. Thus, \eqref{eq:Lyapunov} holds. It follows that 
$(\mathbb{E}(g_{1,n}^2))^{-1/2}I_1$ converges in distribution to a standard normal random variable.

\subsubsection*{Step 2}

Set $\psi_{j+p}=\psi'_j, j=1,\dots, p,$ and denote by $\Psi$ a diagonal matrix of dimension $2p$ with diagonal elements $\psi_j, j=1,\dots, 2p.$ Let also
$\bar\Psi$ be a diagonal matrix of dimension $2p$ with diagonal elements $\bar\psi_j=\sqrt{n^{-1}\sum_{i=1}^n U_{i,j}^2}, j=1,\dots, 2p$. By Assumption \ref{ass:penloadings} we have $\psi_j\ge \bar\psi_j$. Hence,
\begin{equation*}
| I_2 | \le \lVert \Psi (\hat \eta - \eta_0) \rVert_1 \lVert \Psi^{-1} \sqrt{n} \E_n\left[\nabla_{\eta} g(Z,\theta_0, \eta_0)\right] \rVert_\infty
\le \lVert \Psi (\hat \eta - \eta_0) \rVert_1 \lVert \bar\Psi^{-1} \sqrt{n} \E_n\left[\nabla_{\eta} g(Z,\theta_0, \eta_0)\right] \rVert_\infty.
\end{equation*}
Here, the term $ \lVert \bar\Psi^{-1} \sqrt{n} \E_n\left[\nabla_{\eta} g(Z,\theta_0, \eta_0)\right] \rVert_\infty$ is a maximum of self-normalized sums of variables $U_{i,j}$ and it can be bounded by using standard inequalities for self-normalized sums, cf. Lemma  \ref{lem:ModerateDev} below. From the orthogonality conditions and Assumption \ref{CovBound} we have $\mathbb{E}(U_{i,j}) = 0$,  $\mathbb{E}(U_{i,j}^2) \ge c_1$ and $\mathbb{E}(\vert U_{i,j}\vert^3) < \infty$ for any $i$ and any $j$. By Lemma  \ref{lem:ModerateDev} 
and the fact that $\Phi^{-1}(1-a) \leq \sqrt{-2\log(a)}$ for all $a\in (0,1)$ we obtain that, with probability tending to 1 as $n\to\infty$,
\begin{equation*}
\lVert \bar\Psi^{-1} \sqrt{n} \E_n\left[\nabla_{\eta} g(Z,\theta_0, \eta_0)\right] \rVert_\infty \le \Phi^{-1}(1-\gamma/2p) \le \sqrt{2 \log(2p/\gamma)}\lesssim \sqrt{\log(p)}.
\end{equation*}
Next, inequalities \eqref{beta1} and \eqref{ineq:norme1mu1} imply that  $\lVert \Psi (\hat \eta - \eta_0) \rVert_1 \lesssim (s/\kappa_\Sigma) \sqrt{\log(p)/n}$ with probability tending to 1 as $n\to\infty$. Using these facts and the growth condition (i) in Assumption \ref{ass:grow} we conclude that $I_2$ converges to 0 in probability as $n\to\infty$.

\subsubsection*{Step 3} \label{steppe3}

Let ${\bf H} := \E_n\left[\nabla^2_{\eta} g(Z,\theta_0, \tilde\eta)\right]\in \mathbb{R}^{2p\times 2p}$ and let $h_{k,j}$ be the elements of matrix ${\bf H}$. We have
\begin{equation}\label{eq3.1}
| I_3 | \le \frac{\sqrt{n}}{2} \lVert  \hat \eta - \eta_0 \rVert_1^2 \; \max_{1\le k,j\le 2p}|h_{k,j}|.
\end{equation}
We now control the random variable $\max_{1\le k,j\le 2p}|h_{k,j}|$. To do this, we first note that 
\begin{align*}
\frac{\partial^2}{\partial \beta \partial \beta^T} g(Z,\theta,\eta) &= -(1-D) h''(X^T\beta) \left[Y- X^T \mu\right]XX^T, \\
\frac{\partial^2}{\partial \mu \partial \beta^T} g(Z,\theta,\eta) &= \frac{\partial^2}{\partial \beta \partial \mu^T} g(Z,\theta,\eta) = (1-D) h'(X^T\beta) X X^T,\\
\frac{\partial^2}{\partial \mu \partial \mu^T} g(Z,\theta,\eta) &= 0.
\end{align*} 
It follows that
\begin{equation}\label{eq3.2}
\max_{1\le k,j\le 2p} | h_{k,j} | \le \max\left(  \max_{1\le k,j\le p} |\tilde h_{k,j}|,  \max_{1\le k,j\le p} |\bar h_{k,j}|  \right),
\end{equation}
where
$$
\tilde h_{k,j} = \frac{1}{n}\sum_{i=1}^n (1-D_i)  h''(X_i^T \tilde \beta)(Y_i - X_i \tilde \mu)  X_{i,k} X_{i,j},
$$
and
$$
\bar h_{k,j} = \frac{1}{n} \sum_{i=1}^n (1-D_i) h'(X_i^T \tilde \beta) X_{i,k} X_{i,j}.
$$
We now evaluate separately the terms $\max_{1\le k,j\le p} |\tilde h_{k,j}|$ and  $\max_{1\le k,j\le p} |\bar h_{k,j}| $.

Note that $\tilde h_{k,j}$ can be decomposed as 
$$
\tilde h_{k,j} = \tilde h_{k,j,1} + \tilde h_{k,j,2}+ \tilde h_{k,j,3}+ \tilde h_{k,j,4},
$$
where 
\begin{align*}
\tilde h_{k,j,1}&= n^{-1} \sum_{i=1}^n(1-D_i) h''(X_i^T \beta_0)(Y_i - X_i^T \mu_0)X_{i,k} X_{i,j}, \\
\tilde h_{k,j,2} &=  n^{-1} \sum_{i=1}^n(1-D_i) h''(X_i^T \beta_0)X_i^T(\mu_0 - \tilde \mu)X_{i,k} X_{i,j}, \\
\tilde h_{k,j,3} &= n^{-1} \sum_{i=1}^n(1-D_i) (h''(X_i^T \tilde \beta) - h''(X_i^T \beta_0))(Y_i - X_i^T \mu_0)X_{i,k} X_{i,j},\\
\tilde h_{k,j,4} &= n^{-1} \sum_{i=1}^n(1-D_i) (h''(X_i^T \tilde \beta) - h''(X_i^T \beta_0))X_i^T(\mu_0 - \tilde \mu)X_{i,k} X_{i,j}.
\end{align*}
It follows from Assumption \ref{CovBound} that, for all $k,j$, 
\begin{equation}\label{eq3.3}
|\tilde h_{k,j,2}|\le C \|\mu_0 - \hat \mu\|_1.
\end{equation}
Here and in what follows we denote by $C$ positive constants depending only on $K$ that can be different on different appearences. Next, from Assumptions
\ref{CovBound} and \ref{assump:h} (i) we obtain that, for all $k,j$, 
\begin{equation}\label{eq3.4}
|\tilde h_{k,j,3}|\le C \|\beta_0 - \hat \beta\|_1, \quad |\tilde h_{k,j,4}|\le C \|\beta_0 - \hat \beta\|_1 \|\mu_0 - \hat \mu\|_1.
\end{equation}
From \eqref{eq3.3}, \eqref{eq3.4}, Theorem \ref{the:nuisparam} and the growth condition (i) in Assumption \ref{ass:grow} we find that, with probability tending to 1 as $n\to\infty$, 
\begin{equation}\label{eq3.5}
\max_{1\le k,j\le p}  (|\tilde h_{k,j,2}| + |\tilde h_{k,j,3}| + |\tilde h_{k,j,4}|) \lesssim (s/\kappa_\Sigma) \sqrt{\log(p)/n} \lesssim 1.
\end{equation}
Next, again from Assumption \ref{CovBound}, we deduce that $|\mathbb{E}(\tilde h_{k,j,1})|\le C$, while by Hoeffding's inequality
$$
P\big( |\tilde h_{k,j,1} - \mathbb{E}(\tilde h_{k,j,1})| \ge x\big)\le 2\exp(- Cnx^2), \quad \forall x>0.
$$
This and the union bound imply that there exists $C>0$ large enough such that, with probability tending to 1 as $n\to\infty$, 
\begin{equation}\label{eq3.6}
\max_{1\le k,j\le p}  |\tilde h_{k,j,1}| \le C(1+  \sqrt{\log(p)/n}) \lesssim 1.
\end{equation}
Finally, combining \eqref{eq3.5} and \eqref{eq3.6} we obtain that, with probability tending to 1 as $n\to\infty$, 
\begin{equation*}
\max_{1\le k,j\le p}  |\tilde h_{k,j}| \lesssim 1.
\end{equation*}
Quite similarly, we get that with probability tending to 1 as $n\to\infty$,
\begin{equation*}
\max_{1\le k,j\le p}  |\bar h_{k,j}| \lesssim 1.
\end{equation*}
Thus, with probability tending to 1 as $n\to\infty$ we have $\max_{1\le k,j\le 2p} | h_{k,j} | \lesssim 1$. On the other hand,  $\lVert  \hat \eta - \eta_0 \rVert_1 \lesssim (s/\kappa_\Sigma) \sqrt{\log(p)/n}$ with probability tending to 1 as $n\to \infty$ due to Theorem~\ref{the:nuisparam}. Using these facts together with \eqref{eq3.1} and the growth condition (i)  in Assumption \ref{ass:grow} we conclude that $I_3$ tends to 0 in probability as $n\to \infty. $ 

\subsubsection*{Step 4}\label{steppe4}

We now prove that if $\hat \theta \to \theta_0$ in probability then  $\frac{1}{n}\sum_{i=1}^n  g^2(Z_i, \hat \theta, \hat \eta) \to
 \mathbb{E}g^2(Z,\theta_0, \eta_0)$ in probability as $n\to\infty$. We have 
$$
g(Z,\theta,\eta) = [D - (1-D) h(X^T \beta) ] [Y - X^T \mu] - D \theta.
$$
Theorem \ref{the:nuisparam} and the growth condition (i) in Assumption \ref{ass:grow} imply that $\|\beta_0 - \hat \beta\|_1$ is bounded by~1  on an event $\mathcal A_n$ of probability tending to 1 as $n\to \infty$. Using Assumption \ref{CovBound} we deduce that, on the event $\mathcal A_n$, the values $X_i^T\beta_0$ and $X_i^T \hat \beta$ for all $i$ belong to a subset of~$\mathbb R$ of diameter at most~$2K$. On the other hand, Assumption \ref{assump:h} (i) implies that function $h$ is bounded and Lipschitz on any compact subset of~$\mathbb R$. Therefore, using again Assumption \ref{CovBound} we find that on the event $\mathcal A_n$ we have
$|h(X_i^T\beta_0)-h(X_i^T \hat \beta)|\le C\|\beta_0 - \hat \beta\|_1$ for all $i$.
It follows from this remark and from Assumption \ref{CovBound} that, on the event $\mathcal A_n$,   
\begin{align*}
|g(Z_i, \hat \theta, \hat \eta)-g(Z_i,\theta_0, \eta_0)| &\le |\hat \theta-\theta_0| +
|h(X_i^T\beta_0)-h(X_i^T \hat \beta)| |Y - X^T \mu_0| +
  |h(X_i^T \hat \beta)| |X_i^T(\mu_0 - \hat \mu)|
 \\
 & \le |\hat \theta-\theta_0| + C\left(\|\beta_0 - \hat \beta\|_1+\|\mu_0 - \hat \mu\|_1+\|\mu_0 - \hat \mu\|_1\|\beta_0 - \hat \beta\|_1\right):= \zeta_n
\end{align*}
for all $i$. Also note that, due to Assumption \ref{CovBound}, the random variables $g(Z_i,\theta_0, \eta_0)$ are a.s.  uniformly bounded. Thus, using the equality $b^2-a^2=(b-a)^2 + 2a(b-a), \forall a,b \in \mathbb{R},$ we get that, on the event $\mathcal A_n$,   
\begin{align*}
|g^2(Z_i, \hat \theta, \hat \eta)-g^2(Z_i,\theta_0, \eta_0)| &\le  C(\zeta_n^2 + \zeta_n).
\end{align*}
for all $i$.
Using this fact together with Theorem \ref{the:nuisparam}, the growth condition (i) in Assumption \ref{ass:grow} and the convergence 
$\hat \theta \to \theta_0$ in probability we find that $\frac{1}{n}\sum_{i=1}^n  g^2(Z_i, \hat \theta, \hat \eta) - \frac{1}{n}\sum_{i=1}^n  g^2(Z_i,  \theta_0, \eta_0)\to 0$
in probability as $n\to\infty$. We conclude by applying the law of large numbers to the sum $ \frac{1}{n}\sum_{i=1}^n  g^2(Z_i,  \theta_0, \eta_0)$.
\hfill$\square$

\subsection{Proof of Theorem \ref{the:nuisparam}}
\label{app:proof_nuis}

\subsubsection{Proof of \eqref{ineq:norme1}}

Recall that $\hat \beta$ is defined as:
\begin{equation} \label{CalibrationLasso}
\hat \beta \in \underset{\beta \in \mathbb{R}^p}{\arg \min} \left( \frac{1}{n} \sum_{i=1}^n [(1-D_i) H(X_i^T\beta) - D_i X_i^T \beta] + \lambda \sum_{j=1}^p \psi_{j} \vert \beta_j \vert\right)
\end{equation}
with penalty loadings satisfying Assumption $\ref{ass:penloadings}$.  Let  $\Psi_d\in \mathbb{R}^{p\times p}$ be the diagonal matrix with diagonal entries $\psi_{1},\ldots,\psi_{p}$.  Let also
$\bar\Psi_d\in \mathbb{R}^{p\times p}$ be a diagonal matrix  with diagonal entries $\bar\psi_j=\sqrt{n^{-1}\sum_{i=1}^n U_{i,j}^2}, \,j=1,\dots, p$. We denote by $S_0\subseteq \{1,\dots, p\}$ the set of indices of non-zero components of $\beta_0$. By assumption, $\text{Card}(S_0) = \norm{\beta_0}_0 \le s_\beta$.

\subsubsection*{Step 1: Concentration Inequality}

We first bound the sup-norm of the gradient of the objective function using Lemma \ref{lem:ModerateDev}.
Recall that for $1\leq j\leq p$ we have $U_{i,j}=\left[ (1-D_i) h(X_i^T \beta_0)-D_i\right] X_{i,j}$. Set $\mathcal{S}_j := \frac{1}{\sqrt{n}}\sum_{i=1}^n U_{i,j} / \bar\psi_{j}$ and consider the event
\begin{equation*}
\mathcal{B} := \left\{ \frac{1}{n} \,\underset{1 \leq j \leq p }{\max} \left| \sum_{i=1}^n \frac{U_{i,j}}{\bar\psi_{j}} \right| \le \frac{\lambda }{c} \right\}.
\end{equation*}
By construction, the random variables $U_{i,j}$ are i.i.d., $\mathbb{E}(U_{i,j}) = 0$, $\mathbb{E}(U_{i,j}^2) \ge c_1$ and  $\mathbb{E}(\vert U_{i,j} \vert^3) \le  C$ by Assumptions \ref{CovBound} and \ref{assump:h}. Using these remarks,  \ref{ass:penloadings} and Lemma \ref{lem:ModerateDev} we obtain
\begin{align*}
P\left( \mathcal{B}^C \right) &=  P\left( \frac{c}{\sqrt{n}} \ \underset{1 \leq j \leq p }{\max} \vert \mathcal{S}_j \vert > c \Phi^{-1}(1 - \gamma/2p) / \sqrt{n} \right)  \\
&= P\left( \underset{1\le j \le p}{\max}\vert \mathcal{S}_j \vert >  \Phi^{-1}(1 - \gamma/2p) \right) = o(1) \ \text{as} \ n\to \infty.
\end{align*}

\subsubsection*{Step 2: Restricted Eigenvalue condition for the empirical Gram matrix}

The empirical Gram matrix is
$$
\hat \Sigma := \frac{1}{n}\sum_{i=1}^n (1-D_i)X_iX_i^T = \frac{1}{n}\sum_{i=1}^n (1-D_i)^2X_iX_i^T.
$$ 
We also recall the Restricted Eigenvalue (RE) condition  \citep{BRT}. For a non-empty subset $S \subseteq \{1,\dots,p\}$ and $\alpha > 0$, define the set:
\begin{equation}
\mathcal{C}[S,\alpha] := \left\{ v \in \mathbb{R}^p : \lVert v_{S^C}\rVert_1 \leq \alpha \lVert v_{S}\rVert_1 , v\neq 0\right\}
\end{equation}
where $S^C$ stands for the complement of $S$. Then, for given $s\in\{1,\dots,p\}$ and $\alpha > 0$, the matrix $\hat \Sigma$ satisfies the RE($s$, $\alpha$) condition if there exists  $\kappa(\hat\Sigma)>0$ such that 
\begin{equation}\label{re}
 \underset{\substack{S \subseteq \left\{1,\dots,p \right\}:\\ {\rm Card}(S) \le s}}{\min} \ \, \underset{v \in \mathcal{C}[S,\alpha]}{\min} \frac{v^T \hat\Sigma v}{\lVert v_S \rVert_2^2}\ge \kappa(\hat\Sigma) .
\end{equation}
We now use Lemma \ref{lem:rudelson}, stated and proved in Section \ref{sec:auxiliary_lemmas} below. Note that  Assumption \ref{ass:grow} implies \eqref{ass:rud} therein and set $V_i=(1-D_i)X_i$. Then, for any $s\in [1,p/2]$ and  $\alpha>0$, $\hat\Sigma$ satisfies the RE($s$,~$\alpha$)  condition, with $\kappa(\hat \Sigma)=c_*\kappa_\Sigma $ where $c_*\in (0,1)$ is an absolute constant, with probability tending to 1 as $n\to\infty$.

\subsubsection*{Step 3: Basic inequality}

At this step, we prove that with probability tending to 1 as $n\to\infty$, $\hat \beta$ satisfies the following inequality (further called the basic inequality):
\begin{equation} \label{basic}
 \tau (\hat \beta - \beta_0)^T \hat \Sigma (\hat \beta - \beta_0) \leq 2\lambda \left( \lVert \Psi_d \beta_0 \rVert_1 - \lVert \Psi_d \hat \beta \rVert_1 \right) + \frac{2\lambda}{c} \rVert \Psi_d (\hat \beta - \beta_0) \rVert_1, 
\end{equation}
where $\tau>0$ is a constant that does not depend on $n$.

\medskip
By optimality of $\hat \beta$ we have
\begin{equation*}
 \frac{1}{n} \sum_{i=1}^n [\gamma_{\hat \beta} (X_i,D_i) - \gamma_{\beta_0} (X_i,D_i)] \leq \lambda \left( \lVert \Psi_d \beta_0 \rVert_1 - \lVert \Psi_d \hat \beta \rVert_1 \right),
\end{equation*}
where $\gamma_{\beta}(X,D) := (1-D) H(X^T\beta) - D X^T\beta$. Subtracting the inner product of the gradient $\nabla_\beta \gamma_{\beta_0}(X_i,D_i)$ and $\hat \beta - \beta_0$ on both sides we find
\begin{align}
 & \frac{1}{n} \sum_{i=1}^n [\gamma_{\hat \beta} (X_i,D_i) - \gamma_{\beta_0} (X_i,D_i) -\left( (1-D_i)  h(X_i^T \beta_0) - D_i \right) (\hat \beta - \beta_0)^T X_i] \nonumber \\
 &  \qquad \leq \lambda  \left( \lVert \Psi_d \beta_0 \rVert_1 - \lVert \Psi_d \hat \beta \rVert_1 \right) - \frac{1}{n} \sum_{i=1}^n \left( (1-D_i) h(X_i^T \beta_0) - D_i \right) (\hat \beta - \beta_0)^T X_i. \label{eq1:beta}
\end{align}
Using Taylor expansion we get that there exists $0 \leq t \leq 1$ such that
\begin{align*}
&\frac{1}{n} \sum_{i=1}^n \gamma_{\hat \beta} (X_i,D_i) - \gamma_{\beta_0} (X_i,D_i) - \left( (1-D_i) h(X_i^T \beta_0) - D_i\right) (\hat \beta - \beta_0)^T X_i  \\
& \qquad = \frac{1}{2}(\hat \beta - \beta_0)^T \left[ \frac{1}{n} \sum_{i=1}^n (1-D_i) X_i X_i^T h'(X_i^T \tilde \beta)\right] (\hat \beta - \beta_0),
\end{align*}
where $\tilde \beta = t \hat \beta + (1-t) \beta_0$. Plugging  this into \eqref{eq1:beta} and using the facts that  $\vert \sum_{i}a_i b_i \vert \leq \|a\|_1\|b\|_\infty$ and $\psi_j \ge \bar \psi_j$ we get that, on the event $\mathcal{B}$, which occurs with probability tending to 1 as $n\to \infty$:
\begin{align} \label{MVTheorem}
 & \frac{1}{2} (\hat \beta - \beta_0)^T \left[ \frac{1}{n} \sum_{i=1}^n (1-D_i) X_i X_i^T h'(X_i^T \tilde \beta)\right] (\hat \beta - \beta_0) \\
 & \quad \le \lambda \left( \lVert \Psi_d \beta_0 \rVert_1 - \lVert \Psi_d \hat \beta \rVert_1 \right) - \frac{1}{n} \sum_{i=1}^n \left( (1-D_i)  h(X_i^T \beta_0) - D_i \right) (\hat \beta - \beta_0)^T X_i  \nonumber\\
 &   \quad \le \lambda\left( \lVert \Psi_d \beta_0 \rVert_1 - \lVert \Psi_d \hat \beta \rVert_1 \right) +  
 \underset{1 \leq j \leq p }{\max} \left| \frac{1}{n}  \sum_{i=1}^n \frac{X_{i,j}}{\bar\psi_{j}}\left( (1-D_i) h(X_i^T\beta_0) - D_i\right) \right| \, 
 \rVert \Psi_d (\hat \beta - \beta_0) \rVert_1 
 \nonumber\\
 &  \quad \le \lambda \left( \lVert \Psi_d \beta_0 \rVert_1 - \lVert \Psi_d \hat \beta \rVert_1 \right) + \frac{\lambda}{c} \rVert \Psi_d (\hat \beta - \beta_0)\rVert_1.  \nonumber
\end{align}
By Assumption \ref{assump:h} we have $h'>0$, which implies  that the left-hand side of \eqref{MVTheorem} is non-negative. Hence 
we have, under the event $\mathcal{B}$,
$$
0 \leq  \lambda \left( \lVert \Psi_d \beta_0 \rVert_1 - \lVert \Psi_d \hat \beta \rVert_1 \right) + (\lambda/c )\rVert \Psi_d (\hat \beta - \beta_0) \rVert_1.
$$
which implies that 
\begin{align}\label{hatb}
 \lVert \Psi_d \hat \beta \rVert_1 &\leq c_0  \lVert \Psi_d \beta_0 \rVert_1
\end{align}
where $c_0=(c+1)/(c-1)$. By Assumption \ref{ass:penloadings}, we have $\max_j \psi_j \le \max_j \psi_{j,\max} \le \bar{\psi}$ where $\bar{\psi}>0$ 
is a constant that does not depend on $n$. On the other hand,  Assumption \ref{CovBound}(ii) and the fact that the random variables $U_{i,j}^2$ are uniformly bounded implies that $\min_j \psi_j \ge \sqrt{c_1/2}:=\ubar{\psi}$  with probability tending to 1 as $n\to \infty$ (this follows immediately from Hoeffding's inequality, the union bound and the fact that $\log(p)/n\to 0$ due to Assumption \ref{ass:grow}(i)).
These remarks and \eqref{hatb} imply that, with probability tending to 1 as $n\to \infty$,
\begin{align}\label{hatb1}
  \lVert \hat \beta \rVert_1 &\leq c_0 \frac{\bar{\psi}}{\ubar{\psi}}  \lVert \beta_0 \rVert_1.
\end{align}
We now use Assumption \ref{assump:h}(ii). If $\norm{\beta_0}_1 \le c_3$, then  $\lVert \hat \beta \rVert_1 \le c_0(\bar{\psi}/\ubar{\psi})c_3$ with probability tending to 1 as $n\to \infty$, so that $\underset{i=1, \ldots, n}{\min} h'(X_i^T\tilde \beta)\ge h'(-K\max(1,c_0 \frac{\bar{\psi}}{\ubar{\psi}} )c_3) >0$ 
where we have used Assumptions \ref{CovBound}(i) and \ref{assump:h}. Otherwise, given Assumption \ref{assump:h}(ii),  $h'\ge c_2$ on the whole real line so obviously $\underset{i=1, \ldots, n}{\min} h'(X_i^T\tilde \beta) \ge c_2$. It follows that there exists $\tau>0$ that does not depend on $n$ such that, with probability tending to 1 as $n\to \infty$, 
\begin{align}\label{hsigma}
\tau v^T \hat \Sigma v \leq v^T \left[ \frac{1}{n} \sum_{i=1}^n (1-D_i) X_i X_i^T h'(X_i^T \tilde \beta)\right] v, \quad \forall \ v\in \mathbb{R}^p.
\end{align}
Using \eqref{hsigma} with $v=\hat \beta - \beta_0$ and combining it with inequality \eqref{MVTheorem} yields \eqref{basic}.

\subsubsection*{Step 4: Control of the $\ell_1$-error for $\hat \beta$}

We prove that with probability tending to 1 as $n\to\infty$,
\begin{equation}\label{beta1}
\| \Psi_d(\hat \beta - \beta_0)\|_1 \leq \Big(\frac{c}{c-1} \Big)\frac{4\bar\psi^2\lambda s_\beta }{c_*\tau \kappa_\Sigma} , \qquad
\|\hat \beta - \beta_0\|_1 \leq \Big(\frac{c}{c-1} \Big)\frac{4\bar\psi^2\lambda s_\beta }{\ubar{\psi}c_*\tau \kappa_\Sigma}.
\end{equation}
It suffices to prove the first inequality in \eqref{beta1}. The second inequality follows as an immediate consequence. \\

We will use the basic inequality \eqref{basic}. First, we bound $\lVert \Psi_d \beta_0 \rVert_1 -  \lVert \Psi_d \hat \beta \rVert_1$. By the triangular inequality,
\begin{equation*}
\lVert \Psi_d \beta_{0,S_0} \rVert_1 - \lVert \Psi_d \hat \beta_{S_0} \rVert_1 \le \lVert \Psi_d (\beta_{0,S_0}  - \hat \beta_{S_0} ) \rVert_1.
\end{equation*}
Furthermore,
\begin{align*}
\lVert \Psi_d \beta_{0,S_0^C} \rVert_1 - \lVert \Psi_d \hat \beta_{S_0^C} \rVert_1 &=  2 \lVert \Psi_d \beta_{0,S_0^C} \rVert_1 - \lVert \Psi_d \beta_{0,S_0^C} \rVert_1 - \lVert \Psi_d \hat \beta_{S_0^C} \rVert_1\\
&\le 2 \lVert \Psi_d \beta_{0,S_0^C} \rVert_1 - \lVert \Psi_d ( \beta_{0,S_0^C} - \hat \beta_{S_0^C} )\rVert_1\\
&\le  - \lVert \Psi_d ( \beta_{0,S_0^C} - \hat \beta_{S_0^C} )\rVert_1.
\end{align*}
The last inequality follows from the fact that $\lVert \beta_{0,S_0^C} \rVert_1 = 0$. Hence,
\begin{align}
& \lVert \Psi_d \beta_0 \rVert_1 -\lVert \Psi_d \hat \beta \rVert_1 + \frac{1 }{c} \lVert \Psi_d (\hat \beta - \beta_0) \rVert_1 \nonumber \\
\le &  \left( 1 + \frac{1}{c} \right) \lVert \Psi_d (\hat \beta_{S_0} -\beta_{0,S_0})\lVert_1 -   \left( 1 - \frac{1}{c} \right) \lVert \Psi_d (\hat \beta_{S_0^C} -\beta_{0,S_0^C})\lVert_1.\label{spars}
\end{align}
Plugging this result in \eqref{basic} we get,  with probability tending to 1 as $n\to \infty$, 
\begin{align} 
 & (\hat \beta - \beta_0)^T \hat \Sigma (\hat \beta - \beta_0)  \nonumber 
 \\
 \leq & \frac{2\lambda}{\tau}   \left[  \left( 1 + \frac{1}{c} \right) \lVert \Psi_d(\hat \beta_{S_0} -\beta_{0,S_0})\lVert_1 - \left( 1 - \frac{1}{c} \right) \lVert \Psi_d (\hat \beta_{S_0^C} -\beta_{0,S_0^C})\lVert_1 \right], \label{BasicAdvEq}
\end{align}
and thus
\begin{align}\label{beta4}
 (\hat \beta - \beta_0)^T \hat \Sigma (\hat \beta - \beta_0) + \frac{2 \lambda }{\tau} \Big(1 - \frac{1}{c} \Big) \|\Psi_d (\hat \beta - \beta_0)\|_1 
 &
 \leq \frac{4\lambda }{\tau} \|\Psi_d(\hat \beta_{S_0} -\beta_{0,S_0})\|_1
 \\
 & \le \frac{4\lambda  \bar \psi \sqrt{s_\beta}}{\tau} \|\hat \beta_{S_0} -\beta_{0,S_0}\|_2, \nonumber
\end{align}
where we have used the fact that ${\rm Card}(S_0)\le s_\beta$ due to Assumption \ref{ass:dimres}. Recall that $c>1$. From inequality \eqref{BasicAdvEq} and the fact that $(\hat \beta - \beta_0)^T \hat \Sigma (\hat \beta - \beta_0) \ge 0$ we obtain a cone condition  $\Psi_d (\hat \beta - \beta_0) \in \mathcal{C}[S_0,c_0]$ for $\Psi_d (\hat \beta - \beta_0)$, which in turn implies (with probability tending to 1 as $n\to \infty$) a cone condition 
$\hat \beta - \beta_0 \in \mathcal{C}[S_0,c_0 \bar{\psi}/\ubar{\psi} ]$ for $\hat \beta - \beta_0$. Therefore, using \eqref{re} (where we recall that $\kappa(\hat \Sigma)=c_*\kappa_\Sigma $), we obtain that,  with probability tending to 1 as $n\to \infty$, 
\begin{align*}
 (\hat \beta - \beta_0)^T \hat \Sigma (\hat \beta - \beta_0) + \frac{2 \lambda }{\tau} \left(1 - \frac{1}{c} \right) \|\Psi_d (\hat \beta - \beta_0)\|_1 \leq \frac{4\lambda \bar\psi \sqrt{s_\beta} }{\tau}\sqrt{  \frac{(\hat \beta -\beta_0)^T \hat \Sigma (\hat \beta -\beta_0)}{c_*\kappa_\Sigma}}.
\end{align*}
Using here the inequality $ab \le (a^2 + b^2)/2, \forall a,b>0$, we find that, with probability tending to 1 as $n\to \infty$,
\begin{align}\label{beta2}
 \frac{1}{2}(\hat \beta - \beta_0)^T \hat \Sigma (\hat \beta - \beta_0) + \frac{2 \lambda }{\tau} \left(1 - \frac{1}{c} \right) \|\Psi_d (\hat \beta - \beta_0)\|_1 \leq  \frac{8\lambda^{2} \bar\psi^2 s_\beta }{\tau^2 c_*\kappa_\Sigma}.
\end{align}
which implies the first inequality in \eqref{beta1}. Since $\lambda\lesssim \sqrt{\log(p)/n}$ the proof of \eqref{ineq:norme1} is complete. 


\bigskip

\subsubsection{Proof of \eqref{ineq:norme1mu}}

Recall that $\hat{\mu}$ is defined as:
\begin{equation*}
  \hat \mu \in  \underset{\mu\in \mathbb{R}^p}{\arg \min}  \left(\frac{1}{n} \sum_{i=1}^n  (1-D_i) h'(X_i^T \hat \beta)  \left(Y_i - X_i^T \mu\right)^2  + \lambda' \sum_{j=1}^p \psi'_{j} \vert \mu_j \vert\right).
\end{equation*}
Let $\Psi'\in \mathbb{R}^{p\times p}$ denote the diagonal matrix with diagonal entries $\psi'_{1},\ldots,\psi'_{p}$.   We will prove that, with probability tending to 1 as $n\to \infty$,
\begin{equation}\label{ineq:norme1mu1}
\lVert \Psi' (\hat  \mu - \mu_{0})\lVert_1\,\lesssim \frac{s}{\kappa_\Sigma} \sqrt{\frac{\log(p) }{n}}.
\end{equation}
Using an argument analogous to that after \eqref{hatb} we easily get that \eqref{ineq:norme1mu1} implies \eqref{ineq:norme1mu}.



\subsubsection*{Step 1: Concentration inequality}

Define $V_{ij}:= (1-D_i) h'(X_i^T \beta_0) \left[ Y_i - X_i^T \mu_0 \right] X_{i,j}$, $\mathcal{S}'_j := \frac{1}{\sqrt{n}}\sum_{i=1}^n V_{ij} / \psi'_{j}$ and consider the event
\begin{equation*}
\mathcal{B}':= \left\{ \frac{2}{n} \ \underset{1 \leq j \leq p }{\max} \left| \sum_{i=1}^n \frac{V_{ij}}{\psi'_{j}} \right| \le \frac{\lambda'}{c} \right\}.
\end{equation*}
The random variables $V_{ij}, i=1,\dots, n,$ are i.i.d. and  $\mathbb{E}(V_{ij}) = 0$,  $\mathbb{E}(V_{ij}^2) \ge c_1$ and $\mathbb{E}(\vert V_{ij} \vert^3) <C$ for all $i,j$ by Assumptions   \ref{assump:h} and \ref{CovBound}. Using Assumptions \ref{ass:dimres}, \ref{ass:penloadings}, and Lemma \ref{lem:ModerateDev} we obtain
\begin{align*}
P( \mathcal{B}^{'C} ) &=  P\left( \frac{c}{\sqrt{n}} \underset{1 \leq j \leq p }{\max} \vert \mathcal{S}'_j \vert > c \Phi^{-1}(1 - \gamma/2p) / \sqrt{n} \right)  \\
&= P\left( \underset{1\le j \le p}{\max}\vert \mathcal{S}'_j \vert >  \Phi^{-1}(1 - \gamma/2p) \right) =o(1) \quad \text{as} \ n\to \infty.
\end{align*}

\subsubsection*{Step 2: Control of the $\ell_1$-error for $\hat \mu$}

Introduce the notation $\gamma_{\beta,\mu}(Z_i) = (1-D_i) h'(X_i^T\beta)\left( Y_i - X_i^T\mu \right)^2$. It follows from the definition of  $\hat \mu$ that
\begin{align*} 
\frac{1}{n} \sum_{i=1}^n [\gamma_{\hat \beta,\hat \mu}(Z_i) - \gamma_{\hat \beta,\mu_0}(Z_i)] \le \lambda' \left( \lVert \Psi' \mu_0  \rVert_1\ - \lVert \Psi' \hat \mu  \rVert_1\right).
\end{align*}
Here
\begin{align*}
\frac{1}{n} \sum_{i=1}^n [\gamma_{\hat \beta,\hat \mu}(Z_i) - \gamma_{\hat \beta,\mu_0}(Z_i)] &= \left( \hat \mu - \mu_0 \right)^T \left( \frac{1}{n} \sum_{i=1}^n (1-D_i) h'(X_i^T\hat \beta) X_i X_i^T \right) \left( \hat \mu - \mu_0 \right)
\\
 &\quad + \, \frac{2}{n} \sum_{i=1}^n (1-D_i)h'(X_i^T\hat \beta)(Y_i - X_i^T \mu_0) X_i^T ( \mu_0-\hat \mu).
 \end{align*}
 Therefore, 
 \begin{align}
& \left( \hat \mu - \mu_0 \right)^T \left( \frac{1}{n} \sum_{i=1}^n (1-D_i) h'(X_i^T\hat \beta) X_i X_i^T \right) \left( \hat \mu - \mu_0 \right) \nonumber \\
\leq & \lambda' \left( \lVert \Psi' \mu_0  \rVert_1\ - \lVert \Psi' \hat \mu  \rVert_1\right) +   \,\frac{2}{n} \sum_{i=1}^n (1-D_i)h'(X_i^T\beta_0)(Y_i - X_i^T \mu_0) X_i^T (\hat \mu -\mu_0) + R_n, \label{eq:minlossfuncmu}
\end{align}
where
\begin{align*}
R_n &=  \frac{2}{n} \sum_{i=1}^n (1-D_i) \left[ h'(X_i^T\hat \beta ) - h'(X_i^T\beta_0)\right] (Y_i - X_i^T \mu_0 ) X_i^T 
\left(\hat \mu -\mu_0\right)
\\
&=  \frac{2}{n} \sum_{i=1}^n (1-D_i)  (Y_i - X_i^T \mu_0 ) h''(X_i^T\tilde\beta)X_i^T(\hat \beta-\beta_0)  X_i^T 
\left(\hat \mu -\mu_0\right)
\end{align*}
with $\tilde\beta=t\hat\beta +(1-t)\beta_0$ for some $t\in[0,1]$.
Introducing the matrix
$$A= \frac{2}{n} \sum_{i=1}^n (1-D_i)  (Y_i - X_i^T \mu_0) h''(X_i^T\tilde\beta)X_i X_i^T,$$
we can write 
\begin{align*}
R_n &= (\hat \beta-\beta_0)^T A
(  \hat \mu-\mu_0).
\end{align*}
%
%
From \eqref{eq:minlossfuncmu} we deduce that on the event $\mathcal{B}'$ that occurs with probability tending to 1 as $n\to \infty$, 
\begin{align*}
& \left( \hat \mu - \mu_0 \right)^T \left( \frac{1}{n} \sum_{i=1}^n (1-D_i) h'(X_i^T\hat \beta) X_i X_i^T \right) \left( \hat \mu - \mu_0 \right) \\
\leq & \lambda' \left( \lVert \Psi' \mu_0  \rVert_1\ - \lVert \Psi' \hat \mu  \rVert_1\right)+ \frac{\lambda'}{c}\lVert \Psi'( \hat \mu - \mu_0) \rVert_1 
+ (\hat \beta-\beta_0)^T A
(  \hat \mu-\mu_0). 
\end{align*}
We now use \eqref{hsigma} (noticing that $\hat \beta=\tilde \beta$ for $t=1$) to obtain that, with probability tending to 1 as $n\to \infty$,
\begin{align}
\tau \left( \hat \mu - \mu_0 \right)^T \hat \Sigma  \left( \hat \mu - \mu_0 \right)
&\leq \lambda' \left( \lVert \Psi' \mu_0  \rVert_1\ - \lVert \Psi' \hat \mu  \rVert_1\right)
+ \frac{\lambda'}{c}\lVert \Psi'( \hat \mu - \mu_0) \rVert_1 \nonumber \\
 &  \quad
+ (\hat \beta-\beta_0)^T A
(  \hat \mu-\mu_0). \label{eq:2}
\end{align}
Next, observe that, with probability tending to 1 as $n\to \infty$, 
\begin{align}\label{eq:3}
(\hat \beta-\beta_0)^T A(  \hat \mu-\mu_0) - (\tau/2) ( \hat \mu - \mu_0 )^T \hat \Sigma  ( \hat \mu - \mu_0 )
\le C (\hat \beta-\beta_0)^T\hat \Sigma(\hat \beta-\beta_0)
\end{align}
where $C>0$ is a constant that does not depend on $n$. To see this, set $u_i=(\hat \beta-\beta_0)^TX_i$ $v_i=( \hat \mu - \mu_0 )^TX_i$ and $a_i = (Y_i - X_i^T \mu_0) h''(X_i^T\tilde\beta)$. We have
\begin{align*}
(\hat \beta-\beta_0)^T A(  \hat \mu-\mu_0) - (\tau/2) ( \hat \mu - \mu_0 )^T \hat \Sigma  ( \hat \mu - \mu_0 )
& =
\frac{\tau}{2n} \sum_{i=1}^n (1 - D_i)\Big(\frac{4a_i}{\tau} u_iv_i -v_i^2\Big)
\\
 & \le \frac{1}{\tau n} \sum_{i=1}^n (1 - D_i)a_i^2u_i^2.
\end{align*}
This implies \eqref{eq:3} since \eqref{ineq:norme1} and Assumptions \ref{CovBound}(i) and \ref{assump:h} garantee that, with probability tending to 1 as $n\to \infty$,  we have $\max_i|a_i| \le C$ for a constant $C>0$ that does not depend on $n$.

We also note that, due to \eqref{beta2}, with probability tending to 1 as $n\to \infty$, 
\begin{align}\label{eq:4}
(\hat \beta - \beta_0)^T \hat \Sigma (\hat \beta - \beta_0) \lesssim   \frac{\lambda^{2} s_\beta }{\kappa_\Sigma}\lesssim \frac{s_\beta\log(p) }{n\kappa_\Sigma}.
\end{align}
Combining \eqref{eq:2}, \eqref{eq:3} and \eqref{eq:4} we finally get that, with probability tending to 1 as $n\to \infty$, 
\begin{align}\label{eq:5}
\frac{\tau}{2} \left( \hat \mu - \mu_0 \right)^T \hat \Sigma  \left( \hat \mu - \mu_0 \right)
&\leq \lambda' \left( \lVert \Psi' \mu_0  \rVert_1\ - \lVert \Psi' \hat \mu  \rVert_1
+ \frac{1}{c}\lVert \Psi'( \hat \mu - \mu_0) \rVert_1\right) + \frac{\bar c s_\beta\log(p) }{n\kappa_\Sigma},
\end{align}
where $\bar c>0$ is a constant that does not depend on $n$. 
Let $S_1\subseteq \{1,\dots, p\}$ denote the set of indices of non-zero components of $\mu_0$. By assumption, $\text{Card}(S_1) = \norm{\mu_0}_0 \le s_\mu$.
The same argument as in \eqref{spars} (where we replace $ \beta_0, \hat \beta, \Psi_d, S_0$ by $ \mu_0, \hat \mu, \Psi', S_1$, respectively) yields
\begin{align*}
& \lVert \Psi'  \mu_0 \rVert_1 -\lVert \Psi' \hat  \mu \rVert_1 + \frac{1 }{c} \lVert \Psi' (\hat  \mu -  \mu_0) \rVert_1 
\\
\le &  \Big( 1 + \frac{1}{c} \Big) \lVert \Psi' (\hat  \mu_{S_1} - \mu_{0,S_1})\lVert_1 -   \Big( 1 - \frac{1}{c} \Big) \lVert \Psi' (\hat  \mu_{S_1^C} - \mu_{0,S_1^C})\lVert_1.\end{align*}
This and \eqref{eq:5} imply that, with probability tending to 1 as $n\to \infty$, 
\begin{align}\label{eq:6}
&\frac{\tau}{2} \left( \hat \mu - \mu_0 \right)^T \hat \Sigma  \left( \hat \mu - \mu_0 \right) 
+
\lambda'  \Big( 1 - \frac{1}{c} \Big) 
\lVert \Psi' (\hat  \mu - \mu_{0})\lVert_1
\leq 4 \lambda'   \lVert \Psi' (\hat  \mu_{S_1} - \mu_{0,S_1})\lVert_1 + \frac{\bar c s_\beta\log(p) }{n\kappa_\Sigma}, 
\end{align}
where we have used the fact that $c>1$. We now consider two cases:

\begin{enumerate}
\item  $   \lambda' \lVert \Psi' (\hat  \mu_{S_1} - \mu_{0,S_1})\lVert_1\le \displaystyle{\frac{\bar c s_\beta\log(p) }{ n\kappa_\Sigma}}$. In this case, inequality \eqref{eq:6} implies 
$$
\lVert \Psi' (\hat  \mu - \mu_{0})\lVert_1\,\lesssim \frac{s_\beta\log(p) }{\lambda' n\kappa_\Sigma}
$$
and \eqref{ineq:norme1mu1} follows immediately since $\sqrt{\log(p)/n}\lesssim \lambda'$. Consequently, \eqref{ineq:norme1mu} holds in this case.
\item $   \lambda' \lVert \Psi' (\hat  \mu_{S_1} - \mu_{0,S_1})\lVert_1> \displaystyle{\frac{\bar c s_\beta\log(p) }{ n\kappa_\Sigma}}$. Then with probability tending to 1 as $n\to \infty$ we have 
\begin{align*}
&\frac{\tau}{2} \left( \hat \mu - \mu_0 \right)^T \hat \Sigma  \left( \hat \mu - \mu_0 \right) 
+
\lambda'  \Big( 1 - \frac{1}{c} \Big) 
\lVert \Psi' (\hat  \mu - \mu_{0})\lVert_1
\leq 5 \lambda'   \lVert \Psi' (\hat  \mu_{S_1} - \mu_{0,S_1})\lVert_1. 
\end{align*}
This inequality is analogous to \eqref{beta4}. In particular, it implies the cone condition, which now takes the form 
$\lVert \Psi' (\hat \mu_{S_0^C} -\mu_{0,S_0^C})\lVert_1 \leq \alpha \lVert \Psi'(\hat \mu_{S_0} -\mu_{0,S_0})\lVert_1$ with $\alpha=5c/(c-1)$.
Therefore, we can use an argument based on the Restricted Eigenvalue condition, which is completely analogous to that after inequality \eqref{beta4} (we omit the details here). It leads to the following analog of \eqref{beta2}:
\begin{align}\label{eq:7}
&\left( \hat \mu - \mu_0 \right)^T \hat \Sigma  \left( \hat \mu - \mu_0 \right) 
+
C\lambda'  
\lVert \Psi' (\hat  \mu - \mu_{0})\lVert_1
\lesssim  \frac{(\lambda')^{2} s_\mu }{\kappa_\Sigma},
\end{align}
where $C>0$ is a constant that does not depend on $n$. Since $\lambda'\lesssim \sqrt{\log(p)/n}$ we get \eqref{ineq:norme1mu1}. Thus, the proof of \eqref{ineq:norme1mu} is complete.\hfill$\square$
\end{enumerate}

\section{Auxiliary Lemmas} 
\label{sec:auxiliary_lemmas}

 \begin{lemma}[Deviation of maximum of self-normalized sums] \label{lem:ModerateDev}
Consider
 $$
 \mathcal{S}_j := \sum_{i=1}^n U_{i,j} \Big(\sum_{i=1}^n U_{i,j}^2\Big)^{-1/2},
 $$ 
 where $U_{i,j}$ are independent random variables across $i$ with mean zero and for all $i,j$ we have $\mathbb{E}[\vert U_{i,j}\vert^3]\le C_1$, $\mathbb{E}[U_{i,j}^2]\ge C_2$ for some  positive constants $C_1, C_2$ independent of $n$. Let $p=p(n)$ satisfy the condition $\log(p)=o(n^{1/3})$ and let $\gamma=\gamma(n)\in(0,2p)$ be such that $\log(1/\gamma)\lesssim \log(p)$. Then,
\begin{equation*}
P\left( \underset{1\le j \le p}{\max}\vert \mathcal{S}_j \vert >  \Phi^{-1}(1 - \gamma/2p) \right) = \gamma \left( 1+ o(1)\right)
\end{equation*}
as $n\to \infty$.
\end{lemma}
\textbf{Proof.} We use a corollary of a result from \cite{SelfNormalized} given by \citeauthor{BelloniChenChernozhukovHansen2012} (\citeyear{BelloniChenChernozhukovHansen2012}, p.2409), which  in our case can be stated as follows.
Let $\mathcal{S}_j$ and $U_{i,j}$ satisfy the assumptions of the present lemma. If there exist positive numbers $\ell>0$, $\gamma>0$ such that  
\begin{equation} \label{ModerateDevCond}
0 < \Phi^{-1}(1 - \gamma/2p) \leq \frac{C_2^{1/2}}{C_1^{1/3}}\frac{n^{1/6}}{\ell} - 1,
\end{equation}
then,
\begin{equation*}
P\left( \underset{1\le j \le p}{\max}\vert \mathcal{S}_j \vert >  \Phi^{-1}(1 - \gamma/2p) \right) \leq \gamma \left( 1+ \frac{A}{\ell^3}\right),
\end{equation*}
where $A>0$ is an absolute constant. 
\medskip

Now, since $\Phi^{-1}(1 - \gamma/2p) \le \sqrt{2\log(2p/\gamma)}$ and we assume that $\log(1/\gamma)\lesssim \log(p)$ and $\log(p)=o(n^{1/3})$ condition \eqref{ModerateDevCond} is satisfied with $\ell=\ell(n)=(n^{1/3}/\log(p))^{1/4}$ for $n$ large enough. Then $\ell(n)\to \infty$ as $n\to \infty$ and the lemma follows.
\hfill$\square$

\begin{lemma}\label{lem:rudelson}
Let $s\in [1,p/2]$ be an integer and $\alpha>0$. Let $V\in \mathbb{R}^p$ be a random vector such that $\|V\|_\infty\le M<\infty$ (a.s.), and set $\Sigma=\mathbb{E}(VV^T)$. Let $\Sigma$ satisfy \eqref{eq:gram} and 
\begin{equation}\label{ass:rud}
s/\kappa_{\Sigma}= o(p)  \ \ \text{as} \ n\to \infty, \ \text{and} \ \ s\lesssim \displaystyle{\frac{n \kappa^2_{\Sigma}}{\log(p)\log^3(n)}}.
\end{equation}
 Consider  i.i.d. random vectors $V_1,\dots, V_n$ with the same distribution as $V$. 
Then, for all $n$ large enough with probability at least $1-\exp(-C\log(p)\log^3(n))$ where $C>0$ is a constant depending only on $M$ the empirical matrix
$\hat \Sigma =\displaystyle{\frac{1}{n}\sum_{i=1}^n V_iV_i^T}$ 
satisfies the RE($s$,$\alpha$) condition with $\kappa(\hat \Sigma)=c_*\kappa_\Sigma $ where $c_*\in (0,1)$ is an absolute constant.
\end{lemma}

\textbf{Proof.}  We set the parameters of Theorem 22 in \cite{Rudelson} as follows $s_0=s$, $k_0=\alpha$, and due to \eqref{eq:gram} we have, in the notation of that theorem, $K^2(s_0,3k_0, \Sigma^{1/2})\le 1/\kappa_\Sigma$ and $\rho \ge \kappa_\Sigma$. Also note that $\|\Sigma^{1/2}e_j\|_2^2 = \mathbb{E}[(V^Te_j)^2]\le M^2$, where $e_j$ denotes the $j$th canonical basis vector in $\mathbb{R}^p$ (this, in particular, implies that $\kappa_\Sigma\le M^2$). Thus, the value $d$ defined in Theorem 22 in \cite{Rudelson} satisfies $d\lesssim s/\kappa_\Sigma$ and condition $d\le p$ holds true for $n$ large enough due to \eqref{ass:rud}. 
Next, note that condition $n\ge x\log^3(x)$ is satisfied for all $x\le n/\log^3(n)$ and $n\ge 3$, so that the penultimate display formula in Theorem 22 of \cite{Rudelson} can be written as $d \log(p)/ \rho \lesssim n/\log^3(n)$. Given the above bounds for $d$ and $\rho$, we have a sufficient condition for this inequality in the form $s/\kappa^2_\Sigma \lesssim n/\log^3(n)$, which is granted by \eqref{ass:rud}. Thus, all the conditions of Theorem 22 in \cite{Rudelson} are satisfied and we find that, for all $n$ large enough, with probability at least $1-\exp(-C\log(p)\log^3(n))$ where $C>0$ is a constant depending only on $M$ we have
\begin{equation}\label{rud}
 \underset{\substack{S \subseteq \left\{1,\dots,p \right\}:\\ {\rm Card}(S) = s}}{\min} \ \, \underset{v \in \mathcal{C}[S,\alpha]}{\min} \frac{v^T \hat \Sigma v}{\lVert v_S \rVert_2^2}\ge (1-5\delta)\kappa_\Sigma,
\end{equation}
where $\delta\in (0,1/5)$ (remark that there is a typo in Theorem 22 in \cite{Rudelson} that is corrected in \eqref{rud}: the last formula of that theorem should be  $(1-5\delta)\|\Sigma^{1/2} u\|_2 \le \frac{\|Xu\|_2}{\sqrt{n}} \le (1+3\delta)\|\Sigma^{1/2} u\|_2$ where $0<\delta<1/5$ [\cite{Rud2020}]). It remains to note that though at first glance \eqref{rud} differs from \eqref{re} (in \eqref{rud} we have ${\rm Card}(S) = s$ rather than ${\rm Card}(S) \le s$), these two conditions are equivalent. Indeed, as shown in \cite[page 3607]{Bellec}, 
$$
 \underset{{S \subseteq \left\{1,\dots,p \right\}: {\rm Card}(S) \le s}}{\cup}  \big\{ v \in \mathbb{R}^p : \lVert v_{S^C}\rVert_1 \leq \alpha \lVert v_{S}\rVert_1 \big\} = \Big\{ v \in \mathbb{R}^p : \lVert v\rVert_1 \leq (1+\alpha) \sum_{j=1}^s v_j^* \Big\}
$$
where $v_1^*\ge \cdots \ge v_p^*$ denotes a non-increasing rearrangement of $|v_1|, \dots, |v_p|$. On the other hand, 
\begin{align*}
\underset{{S \subseteq \left\{1,\dots,p \right\}: {\rm Card}(S) = s}}{\cup}  \left\{ v \in \mathbb{R}^p : \lVert v_{S^C}\rVert_1 \leq \alpha \lVert v_{S}\rVert_1 \right\} 
&\supseteq 
\left\{ v \in \mathbb{R}^p :  \lVert v_{S_*^C(v)}\rVert_1 \leq \alpha \lVert v_{S_*(v)}\rVert_1 \right\}
\\
&= \Big\{ v \in \mathbb{R}^p : \lVert v\rVert_1 \leq (1+\alpha) \sum_{j=1}^s v_j^* \Big\}
\end{align*}
where $S_*(v)$ is the set of $s$ largest in absolute value components of $v$.
 \hfill$\square$


\end{document}